\documentclass[twoside,twocolumn,9pt]{article}
\usepackage{extsizes}
\usepackage[super,sort&compress,comma]{natbib} 
\usepackage[version=3]{mhchem}
\usepackage[left=1.5cm, right=1.5cm, top=1.785cm, bottom=2.0cm]{geometry}
\usepackage{balance}
\usepackage{mathpazo}
\usepackage{bm}
\usepackage{mathptmx}

\usepackage{sectsty}
\usepackage{graphicx} 
\usepackage{lastpage}
\usepackage[format=plain,justification=justified,singlelinecheck=false,font={stretch=1.125,small,sf},labelfont=bf,labelsep=space]{caption}
\usepackage{float}
\usepackage{fancyhdr}
\usepackage{fnpos}
\usepackage[english]{babel}
\addto{\captionsenglish}{%
  
}
\usepackage{array}
\usepackage{droidsans}
\usepackage{charter}
\usepackage[T1]{fontenc}
\usepackage[usenames,dvipsnames]{xcolor}
\usepackage{setspace}
\usepackage[compact]{titlesec}

\usepackage{tikz}
\usepackage{pgfplots}
\pgfplotsset{compat=1.15}
\usepackage{mathrsfs}
\usetikzlibrary{arrows}

\usepackage{epstopdf}

\usepackage{upgreek}  

\DeclareMathOperator*{\Pressuretype}{P}

\DeclareMathOperator*{\FC}{\boldsymbol{F}^C_i}
\DeclareMathOperator*{\C}{\boldsymbol{C}}

\DeclareMathOperator*{\VIRIAL}{\boldsymbol{\Pressuretype\limits^{\scriptscriptstyle{Virial}}}}
\DeclareMathOperator*{\MOP}{\boldsymbol{\Pressuretype\limits^{\scriptscriptstyle{MOP}}}}
\DeclareMathOperator*{\VA}{{\Pressuretype\limits^{\scriptscriptstyle{V\!A}}}}

\DeclareMathOperator*{\NCER}{{\FC\limits^{\scriptscriptstyle{_{NCER}}}}}
\DeclareMathOperator*{\BORG}{{\FC\limits^{\scriptscriptstyle{_{BORG}}}}}
\DeclareMathOperator*{\FLEK}{{\FC\limits^{\scriptscriptstyle{_{FWF}}}}}
\DeclareMathOperator*{\FCV}{{\FC\limits^{\scriptscriptstyle{_{CV}}}}}
\DeclareMathOperator*{\FDBC}{{\FC\limits^{\scriptscriptstyle{_{FDC}}}}}
\DeclareMathOperator*{\Flin}{{\FC\limits^{\scriptscriptstyle{_{Lin}}}}}

\DeclareMathOperator*{\COCT}{{\C\limits^{\scriptscriptstyle{OCT}}}}
\DeclareMathOperator*{\CCV}{{\C\limits^{\scriptscriptstyle{CV}}}}
\DeclareMathOperator*{\Csurf}{{\C\limits^{\scriptscriptstyle{Surf}}}}

\newcommand{\eq}[1]{ Eq.\ (\ref{#1})}

\definecolor{cream}{RGB}{222,217,201}

\begin{document}

\pagestyle{fancy}
\thispagestyle{plain}
\fancypagestyle{plain}{
\renewcommand{\headrulewidth}{0pt}
}

\makeFNbottom
\makeatletter
\renewcommand\LARGE{\@setfontsize\LARGE{15pt}{17}}
\renewcommand\Large{\@setfontsize\Large{12pt}{14}}
\renewcommand\large{\@setfontsize\large{10pt}{12}}
\renewcommand\footnotesize{\@setfontsize\footnotesize{7pt}{10}}
\makeatother

\renewcommand{\thefootnote}{\fnsymbol{footnote}}
\renewcommand\footnoterule{\vspace*{1pt}%
\color{cream}\hrule width 3.5in height 0.4pt \color{black}\vspace*{5pt}} 
\setcounter{secnumdepth}{5}

\makeatletter 
\renewcommand\@biblabel[1]{#1}            
\renewcommand\@makefntext[1]%
{\noindent\makebox[0pt][r]{\@thefnmark\,}#1}
\makeatother 
\renewcommand{\figurename}{\small{Fig.}~}
\sectionfont{\sffamily\Large}
\subsectionfont{\normalsize}
\subsubsectionfont{\bf}
\setstretch{1.125} 
\setlength{\skip\footins}{0.8cm}
\setlength{\footnotesep}{0.25cm}
\setlength{\jot}{10pt}
\titlespacing*{\section}{0pt}{4pt}{4pt}
\titlespacing*{\subsection}{0pt}{15pt}{1pt}

\fancyfoot{}
\fancyfoot[LO,RE]{\vspace{-7.1pt}\includegraphics[height=9pt]{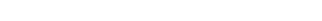}}
\fancyfoot[CO]{\vspace{-7.1pt}\hspace{11.9cm}\includegraphics{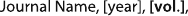}}
\fancyfoot[CE]{\vspace{-7.2pt}\hspace{-13.2cm}\includegraphics{head_foot/RF}}
\fancyfoot[RO]{\footnotesize{\sffamily{1--\pageref{LastPage} ~\textbar  \hspace{2pt}\thepage}}}
\fancyfoot[LE]{\footnotesize{\sffamily{\thepage~\textbar\hspace{4.65cm} 1--\pageref{LastPage}}}}
\fancyhead{}
\renewcommand{\headrulewidth}{0pt} 
\renewcommand{\footrulewidth}{0pt}
\setlength{\arrayrulewidth}{1pt}
\setlength{\columnsep}{6.5mm}
\setlength\bibsep{1pt}

\newcommand{\eg}{${\it e.g.\ }$}
\newcommand{\ie}{${\it i.e.\ }$}

\makeatletter 
\newlength{\figrulesep} 
\setlength{\figrulesep}{0.5\textfloatsep} 

\newcommand{\topfigrule}{\vspace*{-1pt}%
\noindent{\color{cream}\rule[-\figrulesep]{\columnwidth}{1.5pt}} }

\newcommand{\botfigrule}{\vspace*{-2pt}%
\noindent{\color{cream}\rule[\figrulesep]{\columnwidth}{1.5pt}} }

\newcommand{\dblfigrule}{\vspace*{-1pt}%
\noindent{\color{cream}\rule[-\figrulesep]{\textwidth}{1.5pt}} }

\newcommand*{\citen}[1]{
  \begingroup
    \romannumeral-`\x 
    \setcitestyle{numbers}%
    \cite{#1}%
  \endgroup   
}

\makeatother

\twocolumn[
  \begin{@twocolumnfalse}
{\includegraphics[height=30pt]{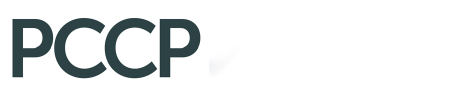}\hfill\raisebox{0pt}[0pt][0pt]{\includegraphics[height=55pt]{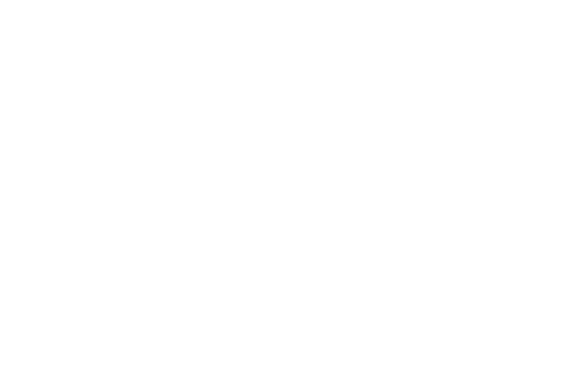}}\\[1ex]
\includegraphics[width=18.5cm]{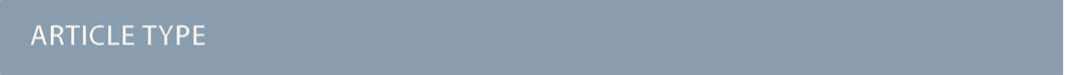}}\par
\vspace{1em}
\sffamily
\begin{tabular}{m{4.5cm} p{13.5cm} }

\includegraphics{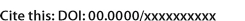} & \noindent\LARGE{\textbf{Multiscale Simulation of Fluids: Coupling Molecular and Continuum }} \\
\vspace{0.3cm} & \vspace{0.3cm} \\

& \noindent\large{Edward R. Smith$^{\ast}$\textit{$^{a}$}} and Panagiotis E. Theodorakis$^{\ast}$\textit{$^{b}$}  \\

\includegraphics{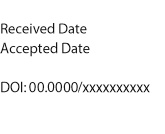} & \noindent\normalsize{
Computer simulation is an important tool for scientific
progress, especially when lab experiments are either extremely costly
and difficult or lack the required resolution. 
However, all of the simulation methods come with limitations. 
In molecular dynamics (MD) simulation, the length and time scales
that can be captured are limited, while computational fluid dynamics (CFD) methods 
are built on a range of assumptions, from the continuum 
hypothesis itself, to a variety of closure assumptions. 
To address these issues, the coupling of different 
methodologies provides a way to 
retain the best of both methods.
Here, we provide a perspective on 
multiscale simulation based on 
the coupling of MD and CFD with each a 
distinct part of the simulation domain.
This style of coupling allows molecular detail
to be present only where it is needed, so CFD
can model larger scales than possible with MD alone.
We present a unified perspective 
of the literature, showing the links between 
state and flux coupling 
and discuss the various assumptions required for both.
A unique challenge in such coupled simulation 
is obtaining averages 
and constraining local parts of a molecular simulation.
We highlight that incorrect localisation
has resulted in an error in the literature for 
both pressure tensor and coupling constraints.
We then finish with some applications, 
focused on the simulation of fluids. 
Thus, we hope to motivate further research in this
exciting area with applications across the spectrum
of scientific disciplines.
} \\

\end{tabular}

 \end{@twocolumnfalse} \vspace{0.6cm}

  ]

\renewcommand*\rmdefault{bch}\normalfont\upshape
\rmfamily
\section*{}
\vspace{-1cm}

\footnotetext{\textit{$^{b}$~Department of Mechanical and Aerospace Engineering, Brunel University London, Uxbridge, Middlesex~UB8~3PH, UK; E-mail: Edward.Smith@brunel.ac.uk }}
\footnotetext{\textit{$^{a}$~Institute of Physics, Polish Academy of Sciences, Al. Lotnik\'{o}w 32/46, 02-668 Warsaw, Poland; E-mail: panos@ifpan.edu.pl}}




Technological advancements in computer software
and hardware, 
combined with scientific ingenuity 
has led to the development of a wealth of 
novel computational methodologies over the years.
This has established computer simulation as a key tool 
in a wide spectrum of fields in science
and engineering across academia and industry.
Moreover, the implementation
of simulation techniques is often provided as
open-source or free software, which allows for the 
widespread use of the methods in various
applications, 
accelerating software development,
and facilitating scientific exchange, validation,
and eventually progress.
As a result, unprecedented perspectives
in scientific research unfold, with 
simulation already having a leading role in the
study of physical and chemical processes,
and novel materials' design.
This is important since simulation
can offer advantages in  cases that lab
experiments are costly, difficult, dangerous, lack
the necessary resolution or are simply impossible.


Luckily, a number of well-established simulation
methods supported by open-source or free software are
available to scientists nowadays. 
However, each simulation method is 
only suitable for capturing a particular 
range of length and time scales of a phenomenon. 
For example, at the scale of most engineering problems,
the continuum assumptions allow a fluid 
to be described using partial differential equations.
These equations require a number of assumptions
to model fluids, such as a constant viscosity 
coefficient \citep{Hansen_et_al}, 
a sharp interface with well-defined surface tension 
\citep{Ghoufi_Patrice_Tildesley}
or a simple relationship between angle and movement 
of a contact line \citep{snoeijerARFM13_review},
which can be shown to break 
down at small enough scales.
At these small scales,
molecular dynamics (MD) simulation is able
to describe the wider range of physics required. 
However, MD is limited to systems in nanometre
length scales and nanosecond time scales. 
With fit-for-purpose hardware and software, MD has
been applied in systems of 
up to 2 billion particles\cite{Phillips2020}
or second time scales.\cite{Mapplebeck2021}
This is still well short of the $10^{25}$ molecules 
present in a single ${\rm m}^3$ of air and, barring a 
revolution in computing power, 
will remain unsuitable for problems 
beyond the microscale.
Hence, it is very much desirable to invent
new simulation protocols that will be able
to combine a multitude of distinct 
methods under the same hood (single simulation), 
thus coherently providing a detailed description
of the system's behaviour across scales.
In turn, this might allow for a better
understanding of the studied phenomena.
The coupling of simulation methods that share
time and length scales allows information
to be easily transferred from one simulation 
technique/domain to the other.
This guarantees a well-defined `interface' between
the two methods/domains.
The basis of this method is 
the exchange of quantities between
the continuum and the MD domain during the simulation
using constraints and averaging to 
guarantee consistency at the interface.

In this perspective article, we aim to 
provide a discussion on the current status
and outlook of coupled simulation approaches with
a focus on the area of fluid dynamics.
Of particular interest is the theoretical 
underpinnings of coupled MD--continuum modelling
in fluids with domain decomposition. 
Some examples of application areas are also given.
Therefore, this does not aim to be a thorough 
review of the literature for
every single simulation method, there have 
been a wide range of reviews. 
\cite{Mohamed_Mohamad, Kalweit_Drikakis11, 
Drikakis_Frank15, bian2020domain, 
Smith_Yates_2018, TONG_2019, Xie_Li23}.
We also do not aim to discuss aspects, such as
the creation of coarse-grained models from bottom-up 
approaches (\textit{e.g.} iterative Boltzmann inversion
methods, free energy methods, \textit{etc.}\cite{Ruhle2009}), 
or the coupling of simulation methods with
experimental data.\cite{Mazurek2021}
Nor is it a discussion of the combination of different force-fields 
in the same simulation method
(\textit{e.g.} MD as in the case of G\={o}EN\cite{Poma2018}
and G\={o}MARTINI models\cite{Poma2017}) used in fields, 
such as biophysics.\cite{Copeland2022,Osaki2022}

Instead, the focus here is on the development of
the theoretical coupling methods
that have matured over the years.
In a 2006 report anticipating the world beyond 2020, 
multi-scale modelling is imagined to be foundational 
for many emerging technologies, shaping 
the future of research.\citep{shapiro2006towards}
This was inspired by quantum to classical coupling using MD,
important to the 2013 Nobel Prize awarded to Arieh Warshe, Michael Levitt, and
Martin Karplus (a brief overview of these methods will be given
in Section~\ref{sec:MDQM}).
However, in many ways the coupling of 
molecular to continuum systems for fluid dynamics
have not taken off in the same way. 
Fluid coupling models remain in their infancy
and very few industrial success stories using 
multi-scale linked simulations 
that incorporate molecular detail have been put out.
Perhaps a major factor is the tragic loss of
two pioneers and champions in multiscale modelling, 
first Jason Reese in 2019 at only 51
then Mark Robbins in 2020 at 64.
The effect on both the scientific community and the research 
funding landscape is profound.
In addition, coupled simulation has never become 
mainstream in the fluid dynamics community,
suffering from a combination of implementational 
complexity and limited or niche areas of application. 
This potential challenge was identified back in 2006
by Ref.~\citen{shapiro2006towards} underlining
the overwhelming software complexity 
requiring ‘industrial-scale’ and industry-wide support.
A limitation also acknowledged succinctly in \citet{TONG_2019}, 
\textit{It is the time to introduce the multiscale methods, especially 
the ‘‘coupling methods”, to the applications on more practical 
multiscale heat transfer and fluid flow problems. Both the 
fundamental and practical researches will benefit from this 
applications, and the multiscale simulations will have a promising future.}
A recent review of multiscale modelling for nanofluids 
\citep{Xie_Li23} suggests further development and improvement 
are required before these methods can be applied to the study of nanofluidics.

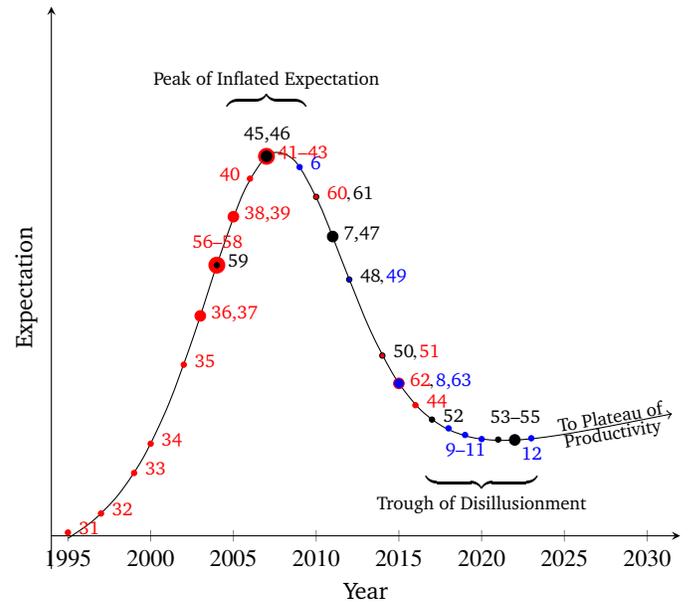
\begin{figure}[bt!]
\begin{tikzpicture}
\begin{axis}[
/pgf/number   format/1000 sep={},
x=2.2mm,y=2.2mm,
axis lines=middle,
ymajorgrids=false,
xmajorgrids=false,
xmin=1994,
xmax=2032,
ymin=-2,
ymax=32,
ytick=false,
ylabel near ticks,
xlabel near ticks,
xlabel=Year,
ylabel=Expectation,]
\draw[color=black, ,domain=1995:2031.5,smooth,->] plot (\x,{-9.3 - (((49.8)/(1 + exp(-0.7*((\x - 2005 - 5)/(1.9))))) 
								              - ((49.8)/(1 + exp(-((0.7*(\x - 2005))/(1.9))))) 
								              - ((23.4)/(1 + exp(-0.08*((\x - 2005 - 5)/(1.9))))))}) ;

\node[anchor=center] at (2007,27) {$\overbrace{\;\;\;\;\;\;\;\;\;\;\;\;\;\;\;}^{\text{Peak of Inflated Expectation}}$};
\node[anchor=center] at (2020,2.5) {$\underbrace{\;\;\;\;\;\;\;\;\;\;\;\;\;\;\;\;\;\;\;\;\;}_{\text{Trough of Disillusionment}}$};
\node[anchor=center, rotate=10] at (2028,7) {${}^{\text{To Plateau of } }$};
\node[anchor=center, rotate=10] at (2028,6) {${}^{\text{Productivity}}$};

\filldraw[red] (1995,0.2) circle (1pt) node[anchor=west]{\citep{OConnell_Thompson}};
\filldraw[red] (1997,1.35) circle (1pt) node[anchor=west]{\citep{Li_et_al}};
\filldraw[red] (1999,3.8) circle (1pt) node[anchor=west]{\citep{Hadjiconstantinou_99}};
\filldraw[red] (2000,5.57) circle (1pt) node[anchor=west]{\citep{Flekkoy_et_al}};
\filldraw[red] (2002,10.35) circle (1pt) node[anchor=west]{\citep{Wagner_et_al}};
\filldraw[red] (2003,13.3) circle (2pt) node[anchor=west]{\citep{Delgado-Buscalioni_Coveney_03_USHER,Hadjiconstantinou_et_al}};
\filldraw[red] (2005,19.3) circle (2pt) node[anchor=west]{\citep{Praprotnik_2005, Werder_et_al}};
\filldraw[red] (2006,21.6) circle (1pt) node[anchor=east]{\citep{DeFabritiis_et_al_06}};
\filldraw[red] (2007,22.95) circle (3pt) node[anchor=west]{\citep{DeFabritiis_et_al_07,Delgado-Buscalioni_DeFabritiis,Kotsalis_et_al}};
\filldraw[red] (2016,7.9) circle (1pt) node[anchor=west]{\citep{KOROTKIN2016446}};

\filldraw[black] (2007,22.95) circle (2pt) node[anchor=south]{\citep{Wang_He,Liu_et_al07}};
\filldraw[black] (2011,18.1) circle (2pt) node[anchor=west]{\citep{Kalweit_Drikakis11, Bugel_et_al11}};
\draw[fill=blue] (2012,15.5) circle (1pt) node[anchor=west]{\citep{Sun_et_al12}${\!}^{,\!}$\textcolor{blue}{\citep{Delgado_Buscalioni_2012}}};
\draw[fill=red] (2014,10.9) circle (1pt) node[anchor=west]{\textcolor{black}{\citep{Zhou_et_al14}}${\!}^{,\!}$\textcolor{red}{\citep{Markesteijn_et_al_2014}}};
\filldraw[black] (2017,7.03) circle (1pt) node[anchor=west]{\citep{Tarasova_2017}};
\filldraw[black] (2021,5.82) circle (1pt) node[anchor=west]{};
\filldraw[black] (2022,5.8) circle (2pt) node[anchor=south]{\citep{Hu_et_al_2019,Papez_Praprotnik_2022,yousefi2022}};

\filldraw[blue] (2009,22.3) circle (1pt) node[anchor=west]{\citep{Mohamed_Mohamad}};
\filldraw[blue] (2018,6.5) circle (1pt) node[anchor=east]{};
\filldraw[blue] (2019,6.1) circle (1pt) node[anchor=north]{\citep{Smith_Yates_2018,TONG_2019,bian2020domain} };
\filldraw[blue] (2020,5.85) circle (1pt) node[anchor=north]{};
\filldraw[blue] (2023,5.9) circle (1pt) node[anchor=north]{\citep{Xie_Li23}};

\filldraw[red] (2004,16.37) circle (3pt) node[anchor=south]{\citep{Flekkoy_DB_Coveney, Delgado-Buscalioni_Coveney_04, Nie_et_al}};
\filldraw[black] (2004,16.37) circle (1pt) node[anchor=west]{\textcolor{black}{\citep{Nie_et_alb}}};
\draw[fill=red] (2010,20.5) circle (1pt) node[anchor=west]{\textcolor{red}{\citep{Borg_et_al}}${\!}^{,\!}$\textcolor{black}{\citep{Sun_et_al10}}};
\draw[fill=blue, draw=red] (2015,9.22) circle (2pt) node[anchor=west]{\textcolor{red}{\citep{Smith_et_al2015}}${\!}^{,\!}$\textcolor{blue}{\citep{Drikakis_Frank15,Delgado-Buscalioni2015}}};

\end{axis}
\end{tikzpicture}
\caption{The literature on domain decomposition coupling as a function of time shown on the Gartner hyper cycle,\citep{Hype_cycle_Gartner} a model for the growth of new technologies where the expectations of a technology over time show a peak followed by a trough. The theoretical contributions are shown in red, the applications in black, and the review articles are shown in blue.}
\label{fgr:Hype}
\end{figure}

The state of the literature is summarised by \citet{TONG_2019}, paraphrasing Fish 
that \textit{‘‘most new technologies began with a native euphoria” when the inventions 
were overpromised. The rapid development led to a ‘‘peak of hype” and followed 
by a period of crash when the immaturity of the ideas was overreacted.} 
This can be seen graphically in Fig. \ref{fgr:Hype}
where the literature on domain decomposition coupling
is shown superimposed on a Gartner hype cycle \citep{Hype_cycle_Gartner}.
Although a fairly arbitrary model for the adoption of a new technology,
it provides a useful perspective on the development of coupling technology.
The literature is coloured by papers developing theory or methods (in red), papers applying 
coupling to problems (black) or publications which review or provide perspectives on coupling (blue).
After a wave of literature developing the method, expectations reach a peak followed
by a period where the important applications and industrial adoption are slow to catch up.
Recent years see a large number of review articles, which characterise the evocatively titled 
`trough of disillusionment', but with some impressive applications starting to appear.
The lack of methodological development (points in red) is also clear after the peak on Fig \ref{fgr:Hype}.
It is for this reason the current perspective focuses on unifying the literature in order to establish a
framework to continue development of the theory to handle the cases needed for increased industrial adoption.
In this perspective we set out the state of the art in theoretical development, before 
listing the recent applications which demonstrate the coupling method, in the hope it can lead the field to
the `plateau of productivity'.
Although a large factor in getting to productivity is the software for coupling, 
this will not be detailed here and have been omitted from Fig \ref{fgr:Hype}.
A range of coupling software exists, 
which can be divided into monolithic \citep{LAMMPS_OPENFOAM},  
frameworks \citep{MUSCLE, MCT} and libraries \citep{Smith_et_al_CPL, MUI, MaMiCo}.
Many of these types of coupling models are summarised in various
published reviews \citep{Taxonomy_Coveney, BORGDORFF2013465, Groen_et_al_2019} 
including previous work by the authors.\citep{Smith2020}
Instead we focus on the theoretical aspects of the problem.

There are a disparate range of methodologies for 
domain decomposition coupling, 
often with researchers working using a preferred model. 
With this in mind, we focus in this perspective on 
trying to present a unified framework which can link together
various models applied throughout the literature.
We show the similarity between the different approaches,
and provide a rigorous underpinning to unify the literature and
provide a tool to enable new theoretical developments.

Through our discussion, 
we aim to 
lower the barrier for embarking on 
hybrid multiscale computer simulations. 
The rapidly evolving
field of machine learning for fluids 
looks set to provide promising
ways of course-graining and coupling
through data-driven approximation 
of underlying physical
processes.
However, such data driven machine
learning methods cannot replace 
the theoretical development of schemes 
which respect conservation laws,
as focused on in this perspective.
Moreover, such mathematical forms for coupled conservation 
laws are likely to be essential ingredients 
to build constraints into 
physics-inspired neural network \cite{Raissi2019}.
This is particularly important, as building the
new generation of powerful coupled (hybrid)
simulation methods and exploiting the 
increasing capabilities in software and hardware
requires scientists with broader computational and
scientific skills and creativity,
based on a deeper knowledge
of multiple simulation methodologies
across scientific fields.
Thus, our perspective article also provides
opportunities for experts that are 
thinking of exploring the capabilities of
hybrid simulation schemes in their 
research areas.

%

\section{Brief Perspective on Other Coupling Approaches}
\label{sec:MDQM}

\subsection{\textit{Ab initio} Molecular Dynamics}
\textit{Ab initio} MD methods
belong to this category of coupled simulation approaches.
Here, classical examples are the Born--Oppenheimer
MD, the Ehrenfest MD, and the Car--Parrinello 
MD (CPMD) methods.\cite{Marx2009,Car1985} 
The goal of these methods is to 
enhance the capability of MD in describing
processes that depend on the electronic
degrees of freedom and, also, better describe
many-body effects.
From the perspective of quantum methods
the use of MD enables a faster evolution of
the nuclear positions, based on empirical
interactions. The aim of \textit{ab initio} MD
is to carry out the classical dynamics of nuclei
and use a quantum dynamics approach for electrons.
Among the \textit{ab initio} MD methods,
the most prominent is probably 
Car--Parrinello MD (CPMD),
since it allows for larger
time steps in the classical approach and
avoids solving the electronic structure problem
at each time. 
The method has been well-established over
the years with further developments that
include its application
in various statistical ensembles.
The electronic
degrees of freedom are introduced into
the model as additional degrees of freedom
as part of an extended Lagrangian equation
of motion, which evolves both the ionic
and electronic degrees of freedom. 
As an advantage, CPMD does not require
matrix diagonalisation, as in the case
of the Born--Oppenheimer MD, and electrons
are assumed to be in the electronic ground state
(adiabatic approximation) during the 
motion of the ions (nuclei). 
The application of the CPMD method in
fluids includes mainly examples such 
as the proton transport in bulk water.\cite{Izvekov2005} Research in
this area is active and new developments
are expected to enable larger time
and length scales while at the same
time holding the capabilities of
describing the electronic state of a
system, which would allow for the study of
more complex phenomena, especially those
relying on charge interactions. \textit{Ab initio}
MD simulations are still more focused
on material design (structure and properties,
including electronic properties), rather than
fluid dynamics problems that require
the simulation of flow changes or even
heat transfer. 
However, subjects such as tribology
require a detailed model of the fluid 
and solid material with some interface interactions
dependant on the \textit{Ab initio} detail.\citep{VAKIS2018169} 
As computing capabilities continuously
increase, a wider range of applications
in fluid dynamics might start to employ
 \textit{ab initio MD} methods
 in the future.
These might be used to parametrise 
intermolecular interactions at these interfaces
for use in classical solvers, or applied directly in coupled QM--MM methods
as described in the next subsection \ref{sec:CQMM}.

\subsection{Coupling Quantum--Molecular Mechanics}
\label{sec:CQMM}
While in the case of \textit{ab initio} MD
methods the classical and electronic degrees
of freedom are attempted to be solved together,
for example, based on an extended Lagrangian of 
motion,\cite{Car1985} a more common strategy
is to use methods with different resolution for 
different parts of the system and establish
the interface region between the methods. 
In this category of coupling fall 
quantum mechanics/molecular mechanics (QMMM)
methods,\cite{Warshel1976} 
which have been very popular in investigating
biological systems.\cite{Brunk2015}
Here, a part of the system is treated 
classically, while another part
that is of particular interest for a process
is simulated quantum-mechanically.
The classical approach usually refers to
molecular-mechanics (MM) simulations, which
include a detailed description of the system,
that is, all atoms are explicitly
modelled, angle and dihedral-angle potentials,
point charges, \textit{etc.}.
A focus of this method is to establish accurate
descriptions of the interactions between the
QM and the MM systems, especially for situations
of covalently bonded atoms and electrostatic
QMMM interactions, or when particles are
moving between the QM and the MM domains. 
Various approaches 
trying to address these issues have been
considered, with this area being under intensive
development, for example,
through various embedding methods, such
as mechanical, electrostatic and polarised, 
and boundary schemes, such as link
atom, boundary atom, and localised-orbital
schemes. 
A common approach for QMMM simulations is
ONIOM,\cite{Dapprich1999} 
which is available in most popular open-source 
MD softwares with a focus on biological
systems. However, most simulations in fluid
dynamics do not require the focus on the
nature of such properties and such methods
are less used in fluids. Still,
as we manage to gain greater ability 
to describe the properties and structure
of fluids and simulate ever larger systems,
the ability of obtaining electronic
structure information can emerge as
an asset in the future, especially in 
micro-chip technologies. Efforts to couple
various methods in this area also remain
in the focus of simulation, such as 
the recent effort to couple 
Density Functional Theory 
with Dynamical Mean Field Theory within
the Framework of Linear Combination of 
Numerical Atomic Orbitals.\cite{Qu2022}
It will be interesting to couple such methods
with larger time- and length-scale resolutions
in the case of fluids simulation. 
In tribological applications, such coupled
methods have recently been applied to
model shear and boundary lubrication 
where reactions would be expected to occur.\citep{RESTUCCIA2020109400}
Finally, method developments in 
the coupling of QM and MM domains 
can provide a basin of ideas for the
molecular--continuum coupling, which
is important in fluid dynamics phenomena
and will be discussed in more detail later.

\subsection{Coupling Molecular-Molecular and Molecular-Mesoscale}
\label{sec:MM-MM}

There are a range of cases where the coupling 
between discrete methods can be beneficial. 
For example, coupling Monte Carlo (MC) and MD 
methods can take place
for both all-atom and coarse-grained models,
and at the molecular and mesoscale level. 
The method can be implemented as a multistep
approach \cite{Gilabert2020} or within 
the single simulation in cycles of MC
and MD,\cite{Chen2021} where MC can refer
to a range of different methods, \textit{e.g.}
Wang--Landau.\cite{Wang2001}
Since both methods are particle-based
in the context of fluids
and the same force-field is used (unless
coupling takes place at the level of 
model description, as, for example, in the 
case of AdResS\cite{Praprotnik2005,Praprotnik2008}), 
the coupling is mainly of a technical nature.
In this approach, the advantages of both
methods are exploited in the same simulation. 
On the one hand, in the case of MC
a large selection
of potential moves are available to facilitate
larger conformation changes of the system
and the exploration of the phase space
of the system including complex moves
which would be unlikely to occur 
even in long MD runs. 
Moreover, MC offers the advantage of
directly sampling the energy of a system.
On the other hand,
in the case of MD the dynamics can be obtained
as a function of time, and the molecular velocities
are available for each system snapshot. In addition,
MD is a method that is easy to scale
on massively parallel computing architectures. 
One aspect that requires consideration is
differences related to the concept of time in the two
methods, as the time in MC is something that
can be only indirectly defined, for example, through
diffusion. 
Regarding applications of MC--MD schemes
in fluids, the approach has been particularly
useful in soft matter systems with complex
molecules and open-source software is generally 
available, which can combine MD and MC in various
ensembles.\cite{Barhaghi2022} An increase
in the computational efficiency has been
noted when configurational bias is included
in the MC scheme. Typically, such simulations
can be beneficial for simulation of molecules,
for example regarding the hydration of a
buried binding
pocket in bovine pancreatic trypsin
inhibitor.\cite{Barhaghi2022} 
Although different bias can be included in
both the MD and the MC approaches, MC allows
for a greater flexibility since MD requires a
more careful treatment of the system dynamics.
Approaches, such as metadynamics,\cite{Laio2002} 
are a possible route for biasing MD simulations
to efficiently explore the phase space 
(characterised by the collective variables) 
of a system allowing the bias to be assessed.
Still such approaches are more mature
in the area of MC simulation, which
may provide further motivation toward
coupled MC--MD simulations.

The coupling of MD with classical density 
functional theory (DFT) can be viewed
as an embedded method, since
information obtained from the MD simulation is
communicated and analysed by using the 
classical DFT level of theory.\cite{Milchev2018,LoVerso2011,Theodorakis2022}
Classical DFT
methods are dedicated to acquiring 
free energy expressions that are suitable for
describing the characteristics of a system.
Apart from the ideal free energy term, 
those expressions depend on the system,
and are different for polymer, colloids, \textit{etc.}
Then, the equilibrium density distribution
is self-consistently obtained through 
an iterative procedure that aims at 
minimising the free energy. The accuracy
of the theoretical assumptions and the
ability to provide analytical expressions for
the same will determine the outcome of
the DFT framework. 
In the case of the MD--DFT approach, one
does not need to self-consistently solve the
equations to obtain the density of the system
at each grid cell. 
The detailed density distribution based on
the particles' position 
is provided by the MD simulation
and the DFT can be used to determine the different
free energy components and their relative
contribution to the free energy, which is
often important for identifying key
physical aspects of the 
system.
A recent example of such an approach is the
application of the method to identify the
free energy term that mostly contributes to
the rugotaxis motion of droplets on wavy
substrates.\cite{Theodorakis2022}
Another similar example
of this kind of coupling, the reference
interaction site model (RISM) can be 
coupled with MD or MC simulations, thus
avoiding the necessity of iteratively
solving the RISM equations, as has been
shown in the calculation of solvation
free energies of several small 
molecules.\cite{Freedman2003} 
In this kind of coupling, 
parts of the theory that are difficult
to obtain analytically, can be provided
as data to the different theories,
as, for example in the case of MD--DFT coupling. 
This aspect is important for
an accurate theoretical description of
a system, but, also, to validate and
improve the theory. In this kind of methods,
MD and MC are equivalent in providing the
necessary data for the coupling.

In another form of coupling, one of the
methods can be at the molecular scale
to readily include the molecular-level
detail of the system
and the other can address the mesoscale
description of the fluid flow. 
Lattice Boltzmann (LB) is often identified
as one such mesoscale method, although
as it is often tuned to reproduce CFD style behaviour,
so is perhaps closer to a continuum method.
The LB origins in the Boltzmann equation still 
qualify it as somewhat more fundamental than CFD,
 opening up potential coupling approaches 
using the single particle distribution functions which are
not possible using continuum models.
Coupling examples using MD--LB allows for exploiting the
ability of MD to deal with the simulation of
molecules, which is important to describe
the molecule--molecule interaction, but
the same time also allows for a hydrodynamics-based
description of the system.\cite{Mackay2013}
Coupling of LB with MC has not been reported
in the literature to the best of our knowledge.
Also, LB can be coupled with particle-based
models through Euler--Lagrange approaches,
and various such examples already exist in
the literature.\cite{Filippova1997} 
Variation on these LB methods\cite{Liu2021} 
are often inspired by method previous developed
in fluid dynamics,
while various approaches can be considered for
the particles, such as the discrete element
method (DEM).\cite{Zhang2021} We will not
expand here our discussion on all possible
particle-based models available in the literature,
since this clearly goes beyond the scope.
Coupling of LB with MD offers advantages as
both methods aim at modelling motion of 
fluids and are both massively parallelised 
and suitable
for a range of diverse system geometries.
The coupling to the LB equations can take place
through an additional local external force
to the equations and interpolation 
protocols, while at the same time the fluid also
acts as a heat bath for the MD particles.
The approach
has been demonstrated for complex 
fluids,\cite{Mackay2013}
and recently for MD particles interacting via
the MARTINI force-field.\cite{Yu2020}
Future directions in this area of coupling
may include the incorporation of long-range
potentials and devising new interpolation 
schemes for the coupling of the LB
and MD domains. This might include
developments on the theoretical descriptions
as well as technical aspects related to the
linking of lattice and off-lattice 
simulation models.

The extension to the simulation of
biological molecules is particularly 
attractive for simulating solvent adequately
far away from the biological molecule
that we are interested to study.\cite{Zavadlav2018}
When this type of coupling is 
also combined with different simulation box
geometries can probably further 
minimise the computational
resources required to simulate the 
solvent surrounding a biomolecule,
as is commonly done in the area of
biophysics, for example, for
simulating proteins in solvents.

Finally, the coupling of molecular-scale models may include a
variety of different particle-based models, including
mesoscale models, such as 
dissipative particle dynamics methods (DPD).
These mesoscopic models can sit as a coupling buffer between 
atomistic and continuum hydrodynamics.\cite{DelgadoBuscalioni2009}

\subsection{Coupling Continuum to Continuum}

In the more general category of coupling in
computational fluid dynamics, 
one can add the techniques
used to add particles to continuum flows or modelling
of fluid--solid interactions combining different approaches,
\textit{e.g.} finite element Analysis (FEA) with CFD.
We could also consider modelling fluid in 
different reference frames with the moving fluid
considered as a particle.
%
%
This is commonly known as 
Euler--Lagrangian simulation approach.
There are a number of 
different models suitable for simulating dispersed 
phases (\textit{e.g.} colloids, droplets, bubbles, sand)
in continuum flows, \textit{i.e.} two-phase or more
generally multiphase flows. 
Rather more common in the literature are
studies that deal with the coupling of 
FEM with DEM\cite{Haddad2016} for investigating
various phenomena of particles in flows,
including heat transfer processes.
An alternative
is the Euler--Euler approach for 
simulating such systems., 
with multiphase effects taken into account in the
fluids properties and through closure relations. 
In contrast, in the case of Euler--Lagrangian schemes, 
the continuous medium can be 
modelled by the continuum equations
(\textit{e.g.} momentum equation), while 
separate equations dictate the motion of the
particles (Lagrangian approach), for 
example, Newton's equation with interactions
between the particles. Coupling
the two systems of equations is the goal of
this approach with various options for
treating the particles being available. 
An example here from recent
work is the simulation of cloud
formation.\cite{Denys2022} In this case, 
`particles' can even
refer to surfactant-laden droplets with
different properties, which can even change
during the simulation as a result of
droplet coalescence.
Moreover, the model incorporates effects
that arise from the reduction of surface tension
due to the presence of surfactants by
adopting a statistical physics (stochastic) approach for
droplet processes (\textit{e.g.} coalescence)
based on the superdroplet method.\cite{shima2009super}
This indicates the variety of possibilities
that can be used for the particle models
in an Euler--Lagrangian approach, including
the coupling of ideas between computational
fluid dynamics and statistical physics,
which significantly extends the range
of applications even when the approach
is solely applied in the macroscale domain itself.
Euler--Lagrangian models
constitute a very active field of research
in computational fluid dynamics, which 
may benefit from some of the techniques described in 
Sec.~\ref{sec:coupmolcon}. 
The application of a particle-based 
mesoscale method,
\textit{i.e.} smoothed-particle
hydrodynamics (SPH) with a continuum approach,
\textit{i.e.} finite element method (FEM) has 
also been reported in the literature.\cite{Zhang2011}
Further work and exchange between the 
particle- and the
continuum-simulation communities
may enable the coupling of continuum models with
a range of particle-based approaches
towards novel applications of 
multiscale simulation. 
%

\section{Coupling Molecular to Continuum}
\label{sec:coupmolcon}

Most of the focus in fluids dynamics 
relates to the motion of fluids and particularly 
on the interactions at the interfaces.\cite{Theodorakis2015,Theodorakis2021b,Theodorakis2017,Theodorakis2022,Kajouri2023}
These interfaces can be between a solid and a liquid,
or where two different fluids meet such as a liquid--vapour coexistence\cite{morgado2016saft}
or between two immiscible liquids.\cite{xu2022bridge}
These regions typically require a detailed molecular picture
due to rapid changes and complexities of the interface itself.
This detail is not required in the bulk, where fluid motion
will be broadly identical and well described by a continuum model. 
As a result, continuum--molecular coupling can be used to put
molecular details only where it is needed.

%
%
%

\subsection{Coupling Types}

\begin{figure}[bt!]
\centering
  \includegraphics[height=5cm]{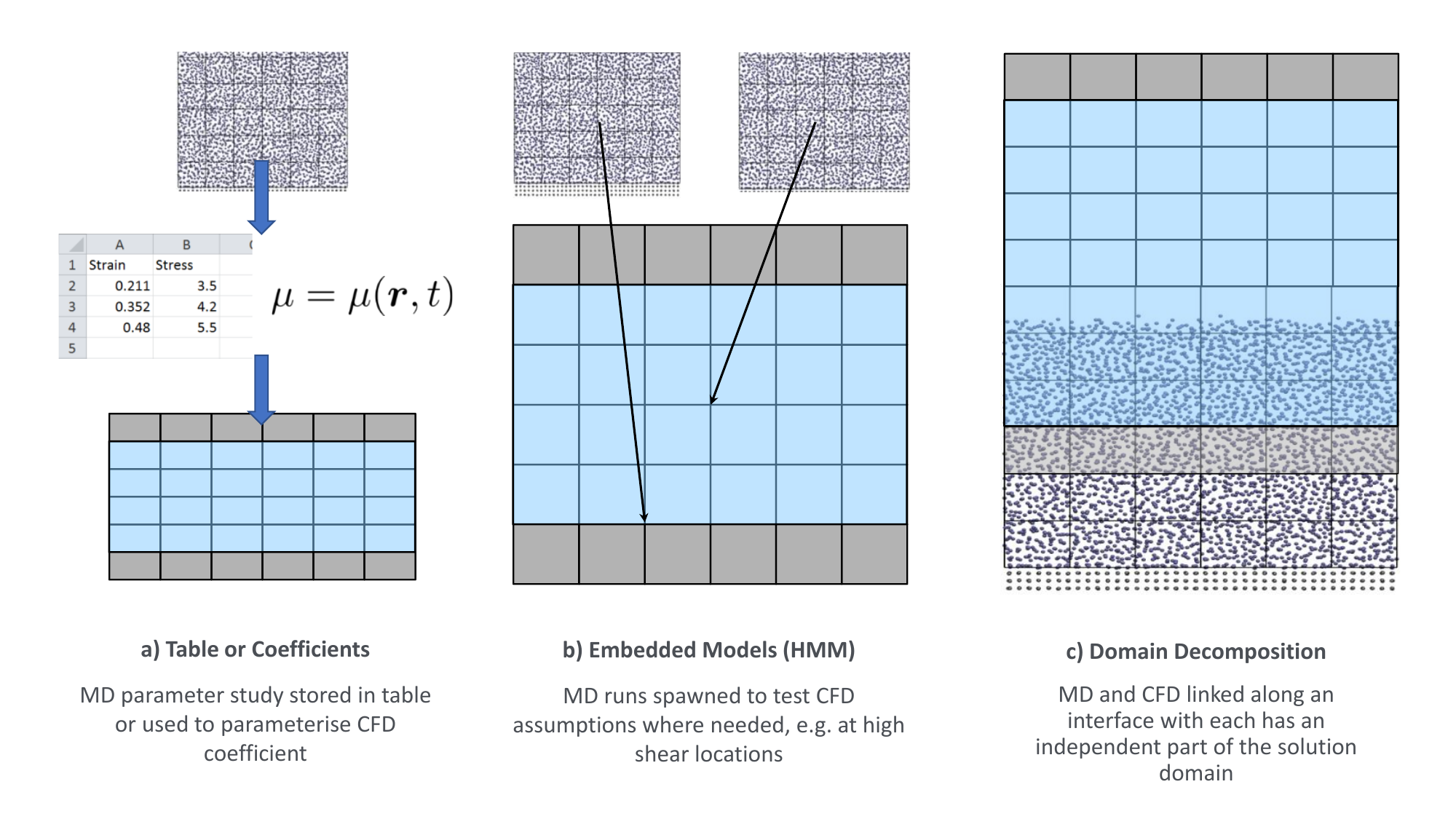}
  \caption{Three categories of classical coupling.}
  \label{fgr:coupling}
\end{figure}

Broadly speaking, classical fluid coupling can be 
divided into three categories \citep{Smith2018} as shown
in Fig.~\ref{fgr:coupling}.
In the simplest example of Fig.~\ref{fgr:coupling}a,
MD is run to obtain parameters for 
computational fluids dynamics (CFD).
This style of parameterisation:
extracting viscosity, 
heat flux or other transport coefficients 
to use in continuum models, is 
the aim of non-equilibrium MD (NEMD)
dating back to the start of molecular simulation.\citep{Evans_Morris, Todd_Daivis_book}
In this type of coupling, both length and time scales
are decoupled so a short small MD run, 
especially using periodic boundaries,\citep{Lees_Edwards}
can be representative of long temporal and spatial scales.
The required assumption is that the MD domain is representative of a larger scale,
accounting for finite-size effects 
\citep{Sandler_Woodcock}
with sufficient averaging time to ensure the
validity of the ergodic hypothesis.
Machine learning could be used here to store more complicated behaviour 
than is possible with constitutive laws, for example using 
Artificial Neural Networks.\cite{Asproulis2013}

Embedded coupling shown in Fig \ref{fgr:coupling}b and described in Ref.~\citen{Ren_07}, 
is also known 
as the heterogeneous multi-scale method (HMM).\citep{Engquist_et_al_2007}
This is used in the case where complexity is too great to be 
characterised by simple coefficients,
so small representative molecular models are run to provide
refinements to the continuum model.
Typically, a state of strain is applied 
to the individual MD runs and the
resulting stress is relayed back to the continuum solver.
Such techniques, \textit{e.g.} SLLOD, \citep{Hoover_SLLOD} 
are valid for one directional shear 
and limited cases of elongation
with use of a coordinate transform.\citep{Frascoli_2006}
This limits their applicability to simple systems
and a general constraint for 
three dimensions is required.\citep{Yamamoto}
HMM has been shown to be a special case 
of the Mori--Zwanzig formalism.\citep{price2018Thesis}
An edited form of this embedded coupling
was given in Ref.~\citen{Borg2013400},
while a detailed discussion of the limitations 
and shortcomings of the HMM method 
was reviewed in Ref.~\citen{Stalter_et_al_2018}.
Embedded coupling runs short and small representative runs
so also decouples time scales from the MD and CFD system.
The required assumption is that these short small runs
reach a steady state, 
which would not change if run for longer.
An example of this is the viscosity, 
where simulation length would only need to be sufficient 
for the molecule state to decorrelate. 
Machine learning again looks to have potential in this field.
This could be built into continuum models, a kind of super resolution sub-sampling
with techniques used for molecular in microscopes \citep{Liu_et_al2019} 
but following similar idea of drawing small scale turbulence into larger simulations 
\citep{fukami_fukagata_taira_2019}, where MD represents a higher-resolution region.

Finally, domain decompositional coupling shown in Fig \ref{fgr:coupling}c uses molecular
detail in a sub-region of the wider domain.
This region is then part of the large continuum simulation
and the two run together with each assigned to its 
respective part of the domain.
This technique is ideally suited to problems
where molecular details are only
required in a local region, 
such as near the wall or at the liquid--vapour interface.
The continuum then becomes a technique for extending
the limited spatial scale possible 
with molecular simulation, 
in that only very small regions of explicit 
MD detail are required.
However, the two simulations are locked into the same 
temporal scale, as the continuum
is then evolving at the same time scale
as the molecular system.
This should therefore be seen as a technique 
for accelerating MD systems, not one for including
molecular detail in continuum scale problems.
However, some techniques for resolving this timescale 
discrepancy do exist.\citep{Liu_et_al, Lockerby_et_al13}
The earliest of these was in the work of \citet{Hadjiconstantinou_et_al}, where
a Schwartz alternating method is used. 
This makes use of the observation that MD
systems often reach a steady state 
quickly for given driving fluxes. 
This quickly converging MD system is 
then iterated with the continuum to obtain
a pseudo-steady solution, which satisfies both.
This technique has been used in \citet{Bugel_et_al11} to coupled interfaces.
In this way, the MD systems can be run 
for short times to provide dynamics equivalent to 
a much longer time step.

%

Given the limitations of domain decomposition style coupling to the molecular time and length scale,
it is reasonable to question why it is worth developing this technique.
Domain decomposition is a technique to accelerate molecular simulation, by expanding the spatial domain that can be simulated using the cheaper CFD model.
It is therefore not a technique which can be used to build molecular detail into CFD, at least in its current form, given the time-scale separation.
The applications areas for this are therefore where MD is essential: the high pressures and strain rates of Tribology;
the initial nucleation of boiling; the chemical reactions at an interface and the fundamental understanding of 
wall--fluid interaction, where direct numerical simulation might be applicable.
Such coupling is the only possible approach in many applications where the continuum assumption or empirically derived constitutive laws fundamentally fail.
The widespread utility of both classical and quantum modelling in materials science shows the potential utility of these non-continuum models.
The promise of the technique for materials justifies the investment to solve these theoretical problems in fluids, and domain decomposition represents the ideal test bed to develop these solutions.
Using the ideas of adaptive grid refinement, a domain decomposition could be imagined to provide insight for just the scales and times it is needed, before being switched off to allow longer time and length scales.


\subsection{Introduction to Domain Decomposition}

\begin{figure}[bt!]
\centering
  \includegraphics[height=5cm]{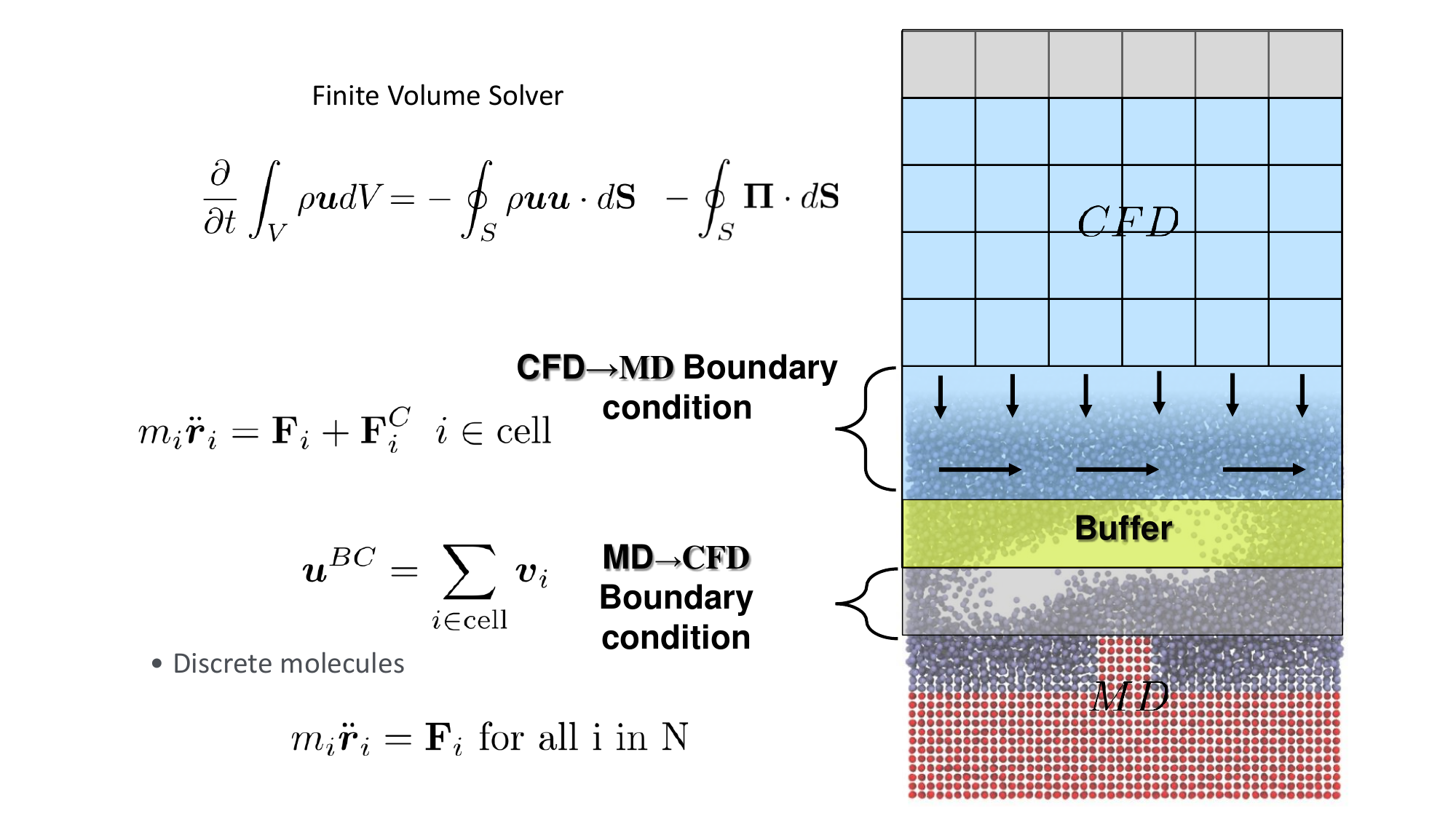}
  \caption{A domain decomposition coupling setup showing 
  1) the averaged region to get the CFD boundary (bottom boundary), 
  2) the region with an applied constraint at the 
  top of the domain where a boundary termination force 
  is applied to stop molecules escaping (top boundary).
  A buffer region is inserted for computational reasons
  so the constrained molecular region doesn't cause 
  a direct feedback with the region averaged 
  to get the boundary. }
  \label{fgr:DD_schematic}
\end{figure}

The anatomy of a domain decomposition coupling 
is shown in Fig.~\ref{fgr:DD_schematic}.
This requires three features, including 
1) a region which is averaged to provide a boundary condition
to the continuum solver,
2) a constrained region where the fluid is driven to
agree with the continuum flow-field and 
3) some method of bounding the MD region at the top,
either using a boundary force, 
a buffer of molecules, or an open boundary
where molecules can be inserted.

The history of domain decomposition coupling for fluid dynamics
starts with \citet{OConnell_Thompson} in 1995.
Despite an initial flurry of interest, work on the theoretical framework, 
especially the constraint forces, largely stopped with the work of \citet{Flekkoy_DB_Coveney}
and has since focused on development of molecular insertion.\citep{bian2020domain}
In comparison, the solid mechanics community has seen extensive research
into the theoretical underpinning of atomistic--continuum 
coupling, traced back to the first papers in the 1970s.\cite{curtin_miller}
This has led to a wide range of different
solid coupling methodologies such as the 
quasicontinuum, CLS method, 
the FEAt, the fully non-local QC (QC-FNL) method 
and the CADD method, all summarised and compared in.\citet{curtin_miller}
Solid mechanics typically uses the finite element form of the
continuum equations, which has a clear mathematical link to 
the continuum equations of motion. 

In solids, lattice deformations are often small so the Cauchy--Born rule can be invoked
to place atoms on finite element nodes and match deformation.
For fluid dynamics, such one to one linking of atoms and nodes is not possible.
Instead, an Eulerian framework is commonly used, tracking the average flow of molecules as they move
through a reference cell or control volume.
The framework to understand these molecular averages comes from statistical mechanics, 
in particular NEMD.
\citep{Todd_Daivis_book, Hoover, Evans_Morris, Tuckermann_book}
However, the NEMD community is distinct from the coupling one, 
perhaps due to the more applied nature 
of coupled simulation.
In this section, we attempt to apply two NEMD techniques to the field of coupling.
In particular linking continuum and molecular equations through \citet{Irving_Kirkwood} 
and constrained dynamics using Gauss' principle of Least Constraint \citep{Hoover, Evans_Morris}
to development of two of the key components of coupling: 
1)  coarse-graining the MD to get continuum fields and 
2) the application of constraints to match MD dynamics to the CFD.


\subsection{Averaging Molecular Systems}
\label{sec:MD_ave}

Domain decomposition coupling requires the molecular system 
be averaged to provide the CFD boundary condition. 
In the literature, coupled boundary exchange
is split into state coupling,\cite{OConnell_Thompson} 
obtaining the velocity and scalar pressure from MD 
simulation, and flux coupling, which directly couples
the stress tensor from the
MD system.\citep{Flekkoy_et_al, Delgado-Buscalioni_Coveney_04}
The Navier--Stokes equation is derived from a stress tensor,
making assumptions about isotropy of the fluid,
Stokes hypothesis, and incompressibility to express 
everything in terms of velocity and pressure. 
As a result, the stress-tensor approach is more general, 
making none of the assumptions but potentially introducing 
more noise into the CFD solver.\citep{Hadjiconstantinou_thesis}
Although pressure measurements are said to be prohibitively
noisy for 
coupling,\cite{Hadjiconstantinou_et_al, Hadjiconstantinou_05}
it can be shown that this depends 
on the definition of error and statepoint of 
the system.\citep{Smith_Thesis}
As a result, the choice should be based on a method which
best ensures conservation laws are valid during coupling.
The finite volume (FV) method is the naturally conservative 
form of the continuum equations, 
using the equation shown in Fig.~\ref{fgr:DD_schematic}
expressed in terms of surface fluxes.
To explore this, in this section we will express both state and 
flux coupling in terms of an explicitly localisation operator, which
allows an equivalent to the FV form to be obtained in the MD system.
This in turn provides a more rigorous expression of the averaging 
operation, which can be used in constrained dynamics.
This operator also allows a form of stress tensor, which 
improves on the virial form commonly used in the coupling literature.

Velocity (state) coupling has traditionally been expressed
in the literature as obtaining the CFD boundary from a
restricted sum over molecules in the MD--CFD overlapping cells,
\begin{align}
 \boldsymbol{u} = \displaystyle\sum_{i=1}^{N_I}  \boldsymbol{\dot{r}}_i = \displaystyle\sum_{i \in \text{cell}}  \boldsymbol{\dot{r}}_i ,
\label{restricted_sum}
\end{align}
This is a well-established binning operation in the MD literature,
and as such has received minimal scrutiny in the development of coupling.
There is, however, a subtlety in Eq.~\ref{restricted_sum},
where the second sum over $N_I$ 
could equally denote molecules at a point $N_I =N_I(\boldsymbol{r}, t) $  
or following a collection of molecules evolving in space,
\textit{i.e.} $N_I =N_I(t) $. 
The set notation $i \in $ cell is more explicit,
clearly stating only molecules located 
inside a cell at a given time.
However, this does not tell us how these
sums should behave when using the calculus, 
for example the time derivative of the sum of molecules $i \in $ cell
must consider how the set itself changes in time.
This seemingly minor consideration 
means the typically used constraints 
developed in the coupling literature 
are missing a critical term, 
as will be discussed in Sec.~\ref{sec:Constraints}.
In continuum fluid mechanics the relationship between 
following a moving collection of fluid particles
and monitoring flow through a fixed region
in space is given by Reynold's transport theorem, 
a central concept in fluid mechanics.\citep{Potter_Wiggert}
To get the molecular equivalent of Reynold's transport theorem, we formalise the 
localisation using a control volume integral of the \citet{Irving_Kirkwood} 
Dirac delta function.\citep{Smith_et_al}
The coarse-grained density and momentum in a control volume
can therefore be written,
\begin{align}
\int_V \rho  dV = \displaystyle\sum_{i=1}^N m_i \vartheta_i = \Delta V \rho^{MD} = M_I \nonumber \\
\int_V \rho \boldsymbol{u} dV = \displaystyle\sum_{i=1}^N m_i \boldsymbol{\dot{r}} \vartheta_i  = \Delta V [\rho \boldsymbol{u}]^{MD}  
\label{CV}
\end{align}
where $\vartheta_i$ is zero outside a given volume 
and one inside, a function comprised from the product of 
Heaviside functions to select molecules inside the Heavisides. 
In the cubiodal case
$\vartheta_i = \Lambda_{xi}\Lambda_{yi}\Lambda_{zi}$ 
where $\Lambda_{\alpha i} = H(\alpha^+ - \alpha_i) - H(\alpha^- - \alpha_i)$ and $\alpha \in \{x, y, z \}$.
The cuboid has volume $\Delta  V$ which is between the limits denoted by superscript $+$ and $-$, 
and can be made to correspond to an identical sized CFD region. 
The notation for the average density $ \rho^{MD}$ 
and momentum $ [\rho \boldsymbol{u}]^{MD} $ inside 
the volume $V$ is introduced and so the average control volume
velocity can be defined as 
$\boldsymbol{u}^{MD} =  [\rho \boldsymbol{u}]^{MD}  /\rho^{MD}$.
This has the advantage that the $\vartheta_i$ function 
takes care of localisation during mathematical operations.
It is for this reason that taking the time evolution 
of \eq{CV} yields the molecular
version of Reynold's transport theorem.\citep{Smith_et_al}
The molecular equations of \eq{CV} are expressed in the 
same form as the mass and momentum used in the finite volume (FV) method.
The FV form is most natural for CFD simulation, owing to
the conservative nature and ease of meshing for
arbitrary geometry.\citep{Hirsch}

Explicit localisation can also be used to derive 
a local stress tensor, to be used in flux coupling.
Here, the volume average (VA) form of pressure is given by an 
integral of the \citet{Irving_Kirkwood} stress tensor,
\begin{align}
\int_V \boldsymbol{P}(\boldsymbol{r},t) dV = \VA \Delta V  = \displaystyle\sum_{i=1}^N m_i \dot{\boldsymbol{r}_i} \dot{\boldsymbol{r}}_i \vartheta_i +  \displaystyle\sum_{i,j}^N \boldsymbol{f}_{ij}  \boldsymbol{r}_{ij} \ell_{ij},
\label{VA_pressure}
\end{align}
and the $\ell_{ij}$ function takes the length of 
inter-molecular interaction inside a volume of 
size $\Delta V$.
For simplicity of presentation, this work doesn't separate the
convective term, \textit{i.e.} we do not do the usual
decomposition $\sum m_i \dot{\boldsymbol{r}_i} \dot{\boldsymbol{r}}_i  = \rho \boldsymbol{u} \boldsymbol{u} +  \sum m_i {\boldsymbol{v}_i} {\boldsymbol{v}}_i  $ where $\boldsymbol{v}_i = \dot{\boldsymbol{r}}_i - \boldsymbol{u}$. 
We have also assumed a linear path of interactions between molecules \citep{Cormier_et_al} to avoid any ambiguity in the definition of pressure.\citep{Schofield_Henderson}
This VA pressure of \eq{VA_pressure} appears similar to the virial pressure,
\begin{align}
\VIRIAL  \Delta V  = \displaystyle\sum_{i=1}^N  m_i \dot{\boldsymbol{r}}_i \dot{\boldsymbol{r}}_i  \vartheta_i +  \displaystyle\sum_{i,j}^N\boldsymbol{f}_{ij}  \boldsymbol{r}_{ij} \vartheta_i.
\label{virial}
\end{align}
However, the pressure of \eq{VA_pressure} satisfies momentum conservation
near a wall, while the virial pressure does not.\citep{Kaihang_et_al2023}
This is because the virial pressure assumes a homogenous system, 
representing a truncation to the first term in an expansion of 
Eq. \ref{VA_pressure}.\citep{Evans_Morris}
Using the virial in a heterogenous MD system 
is known to give spurious pressure peaks
near an interface.\citep{Todd_et_al_95}
Despite this error, the virial pressure is the default in two widely used
open-source codes, LAMMPS and GROMACS, 
at the time of writing \citep{Kaihang_et_al2023} 
and has been widely used in the coupling literature.
As coupled simulations are heterogenous by construction, 
using the virial pressure for coupled simulations will 
very likely be incorrect.

The VA form is an improvement on the virial pressure, but the most 
natural framework for fluid dynamics is the control volume,
or finite volume (FV) form, where conservation
is ensured as fluxes leaving one cell are exactly equal to the fluxes 
into a connected cell.
For a CFD solver in FV form,
the boundary condition is therefore required to be a flux.
In the coupling literature this virial pressure 
is dotted with the surface 
normal $\VIRIAL \cdot \textbf{n}$, to get it as a surface flux.
Often the pressure uses an interpolation operation
with the adjacent continuum cell to get this pressure at the location of the cells surface.\citep{DeFabritiis_et_al_06}
However, a formal version of surface flux already exists in the 
NEMD literature, 
known as the Method of Planes (MOP) pressure.\citep{Todd_et_al_95} 
This is obtained from the flow of momentum carried by 
molecules over a given surface, 
for example take the $x^+$ surface, 
and the intermolecular forces acting over that surface 
\begin{align}
\MOP{}_{\!\! x^+}  & = \MOP{\!\!}_{\!\! x^+}^K + \MOP{\!\!}_{\!\! x^+}^{C}  
\label{total_MOP} 
\end{align}
with the dashes defining surface crossings following the 
notation from the literature.\citep{Flekkoy_DB_Coveney}
\begin{align}
\MOP{\!\!}_{\!\! x^+}^K \Delta A_{x^+} &= \frac{1}{\Delta t} \displaystyle\sum_{i=1}^N m_i \dot{\boldsymbol{r}}_i \dot{x}_i  dS(x_i^t, x_i^{t + \Delta t}) \equiv \displaystyle\sum_{i^{\prime}}^{N_{x^+}}  m_{i^{\prime}} \dot{\boldsymbol{r}}_{i^{\prime}}   \label{Kinetic_MOP} \\
\MOP{\!\!}_{\!\! x^+}^{C} \Delta A_{x^+}  & = \displaystyle\sum_{i=1}^N  \displaystyle\sum_{j \ne i}^N \boldsymbol{f}_{ij} dS(x_i, x_j) \equiv \displaystyle\sum_{ij^{\prime}}^{N_{x^+}} \boldsymbol{f}_{ij^{\prime}}  
\label{Config_MOP} 
\end{align}
These surfaces crossings are exactly defined in terms of 
rigorous mathematical functions obtained from derivatives of $\vartheta_i$.\citep{Smith_et_al, Smith_22}
Here, the crossing function is non-zero 
only when a molecule is crossing a surface of the finite volume,
\begin{align}
 dS(x_s, x_e) = \left[ sign(x^+\!\!\!-x_e) - sign(x^+\!\!\!-x_s) \right]  \Lambda_{yc}\Lambda_{zc}
\label{dS}
\end{align}
where $x_s$ is the start of a straight line 
and $x_e$ is the end, which can represent a molecule $i$ 
evolving in time from position $ x_i^{t} = x_i(t)$ to 
position $ x_i^{t + \Delta t} = x_i(t + \Delta t)$ or
the line of interaction between two molecules $x_i$ and $x_j$.
Note, we have written the kinetic term of \eq{Kinetic_MOP} as the 
integral over a time step 
$(1/ \Delta t)\int_{t}^{t+\Delta t} \delta(x^+ \!\!\! -x_i) \Lambda_{yi}\Lambda_{zi}  dt$,
so it is in the same form as the configurational term.
The expression in \eq{dS} can be directly implemented in 
code, where the signum functions determine whether
the particle has crossed a plane and, for cubic volumes,
a trivial plane-line intersect calculation can give 
the position of crossing $y_c$ and $z_c$.
The crossings are used in 
$\Lambda_{yc}$ and $\Lambda_{zc}$, respectively,
which determine which control volume face it has crossed. 
More generally for complicated control volumes this is 
a ray-tracing problem over every 
bounding volume surface. \citep{Smith_22}
The importance of using crossings was recognised in
\citet{Donev_et_al10}, who used a ray-tracing approach,
essentially equivalent to the kinetic part of the pressure
 \eq{Kinetic_MOP}. 
This was used in coupling between CFD and
Direct Simulation Monte Carlo (DSMC), 
building on earlier work.\citep{garcia1999adaptive}
Using this surface flux form here extends the same approach to 
dense fluid MD simulation, so includes the configurational term.
The MOP form of pressure does not introduce the spurious 
oscillations, which plague the virial form
of \eq{virial} and can be shown to be equivalent to 
the VA form \citep{Heyes_et_al_12} in the limiting case
that the volumes thickness tends to zero.

Most importantly for coupling, the surface 
pressure of \eq{total_MOP}  is the only form that 
guarantees finite-volume
style conservation \citep{Kaihang_et_al2023} 
to machine precision in the MD system,
\begin{align}
\frac{d}{dt}\displaystyle\sum_{n=1}^N m_n \dot{\textbf{r}}_n \vartheta_n 
& =\displaystyle\sum_{\alpha=1}^{Nsurf}  \MOP{}_{\alpha} \Delta A_{\alpha} 
=\sum_{\alpha=1}^{Nsurf}  \int_{A_{\alpha}} \boldsymbol{P} \cdot d \textbf{A}_{\alpha}.
\label{mom_eqn}
\end{align}
The sum is over all surfaces of any bounding volume
and the equality to the continuum form of surface flux
over $N_{surf}$ surfaces of the control volume 
$\sum_{\alpha=1}^{Nsurf}  \int_{A_{\alpha}} \boldsymbol{P} \cdot d \textbf{A}_{\alpha}$ follows directly
from the time evolution of \eq{CV}.
In this way, it behaves in an identical way to the FV form used in CFD, where
ensuring conservation is used to evolve the system in time.
In the next section, the localised momentum and pressure 
presented here are used in the constrained dynamics equations
to derive rigorous localised algorithms. 
By simplifying these we can explore the link between the 
various constraint equations 
used in the literature and provide a general framework to
understand different coupling methodologies.
These forms of constraint can also be expressed in terms 
of exactly conservative finite volumes,
the form used in fluid dynamics solvers.

\subsection{Constraint Force}
\label{sec:Constraints}

A constraint force is a non-unique problem 
in which the total momentum of multiple molecules 
must be driven to some setpoint value.
This setpoint is the momentum in the 
overlapping continuum cells, labelled 
$CFD \to MD$ in Fig.~\ref{fgr:DD_schematic}.
We start by outlining some methods for doing this
based on Maxwell's Demon including particle velocity selection 
and selectively permeable membranes, before discussing 
blending functions inspired by two-phase models from CFD.
We then move on to a presentation of the constraint 
algorithms, which are derived from the minimisation 
principles of physics.
In particular, we show a term is missing from the most 
commonly used expression in the literature,
due to a lack of explicit localisation of the 
form introduced in the previous section.

\subsubsection{Maxwell's Demon}

\citet{Hadjiconstantinou_thesis} applied the transfer 
from the continuum-to-molecular by selecting velocities 
from a Maxwell Boltzmann distribution,
 \begin{align}
 f(\dot{\boldsymbol{r}}) =  \left( \frac{m}{2\pi k_B T} \right)^\frac{3}{2} exp \left( -\frac{m(\dot{\boldsymbol{r}}-\boldsymbol{u})^2}{2k_BT}\right),
 \label{Maxwell_Boltzmann}
 \end{align}
for molecules located near the boundary of the domain. 
Here, $k_B$ is Boltzmann's constant, 
$\boldsymbol{u}$ the continuum velocity,
and $T$ the continuum temperature. 
The molecular domain of interest is surrounded by a 
molecular reservoir. 
The molecule velocities are completely re-defined
in line with the velocities of the overlapping 
continuum region. 
A Taylor series expansion to first order is used to 
establish the velocities and temperatures
to be specified in the 
Maxwell Boltzmann distribution of \eq{Maxwell_Boltzmann}. 
The effects on the dynamics of this `Maxwell's Demon' 
approach are localised near the simulation boundary
and the performance is said to compare favourably to 
constrained dynamics 
approaches.\citep{Hadjiconstantinou_thesis}
The application of the Maxwell Boltzmann distribution 
was later found to result 
in slip,\cite{Hadjiconstantinou_05}
which was reduced by replacing the Maxwell Boltzmann 
distribution function by a non-equilibrium distribution
from the Chapman Enskog expansion 
or previous MD simulations.\citep{Hadjiconstantinou_05} 

\citet{Liu_et_al} introduce a control-style algorithm,
which minimises the disturbance to a system
and avoids applying any forces. 
This is motivated by the observation that 
any applied forces can have undesirable consequences 
as they add energy,
have magnitudes $10^{12}$ times that of gravity,
and assume a constant pressure difference.\citep{Liu_et_al} 
To avoid applying forces,
\citet{Liu_et_al} use a selectively permeable
membrane to bias flow in a certain direction. 
This membrane is also like Maxwell's demon,
effectively reflecting certain molecules 
and allowing others through in a manner that ensures
the required flow profile.

\subsubsection{State Coupling}

State coupling aims to control the state of the system,
namely the density, velocity and temperature as opposed to the 
fluxes of these quantities such as pressure and heat flux.
A set of coupling constraint equations is 
put forward in the original work 
of \citet{Markesteijn_et_al_2014} 
and extended in \citet{KOROTKIN2016446}
In this approach, blending functions are used, 
which are inspired by two-phase flows in hydrodynamics, 
and share similarities with AdResS for molecular
insertions (see section \ref{sec:insertion}) and some of the coupling techniques 
developed in the solid mechanics literature.
\begin{subequations}
\begin{eqnarray}
 \!\!\! \!\!\! \!\!\! \!\!\!	\boldsymbol{\dot{r}_{ i}} & =& \!\!\! \frac{\boldsymbol{{p}_{ i}}}{m_i}+ s \left(\overline{\boldsymbol{u}} - \frac{\boldsymbol{{p}_{ i}}}{m} \right) + \frac{s(1-s) \alpha }{\rho^{MD}} \phi_{\rho} 	\label{Karabasov} \\ 
  \!\!\!\!\!\! \!\!\! \!\!\!\boldsymbol{\dot{{p}}_{ i}} &=& \!\!\! (1-s) \boldsymbol{F}_i +  \frac{s(1-s)}{\rho^{MD}}\frac{\partial}{\partial \boldsymbol{r}} \cdot   \bigg( \alpha \frac{[\rho \boldsymbol{u}]^{MD}}{\rho^{MD}} \phi_{\rho} + \beta \boldsymbol{\phi}_{u} \bigg). \label{Karabasov2}
\end{eqnarray}
\end{subequations}
Here, the flux of density $\phi_{\rho} = \frac{\partial}{\partial \boldsymbol{r}} \left( \overline{\rho} - \rho^{MD} \right)$ and flux of momenta $\boldsymbol{\phi}_{u} = \frac{\partial}{\partial \boldsymbol{r}} \left( \overline{\rho} \overline{\boldsymbol{u}} - [\rho \boldsymbol{u}]^{MD}   \right) $ are introduced where $\rho^{MD}$ and $ [\rho \boldsymbol{u}]^{MD}$ are as defined 
in Eq.~\ref{CV} and the overbar quantities are the 
weighted average of continuum and MD systems, with density $\overline{\rho} = s \rho -  (1-s) \rho^{MD} $ 
and velocity $\overline{\boldsymbol{u}} = s (\rho \boldsymbol{u} -  (1-s)[\rho \boldsymbol{u}]^{MD} )/\overline{\rho} $.
For the case when $s=0$, we reclaim the MD equations, 
$\boldsymbol{\dot{r}_{ i}}=\boldsymbol{\dot{p}_{ i}}/m_i$ 
and $\boldsymbol{\dot{{p}}_{ i}} = \boldsymbol{\dot{{F}}_{ i}} $. 
For $s=1$, the equations becomes, 
$\boldsymbol{\dot{r}_{ i}} = \boldsymbol{u}$  
and $\boldsymbol{\dot{{p}}_{ i}} = 0 $, 
so the molecules are frozen, unaffected by 
intermolecular forces and all moving at 
the velocity of the overlapping continuum.
In the gradual transitions from $s=0$ to $s=1$,
any difference between the density and momentum 
in the two systems act to force the molecules 
in the direction of that difference, 
an example of $s=0.5$ is included in the appendix.

These schemes represent quite a strict constraint, being exact velocity specification
of every molecule in the continuum region.
In some ways, they appear to mix state and flux coupling, 
but the fluxes are included to minimise the difference in density and velocity states between the two systems, 
chosen by trial and error from a choice of soft and hard constraints.\citep{Markesteijn_et_al_2014}
The forcing of \eq{Karabasov} and \eq{Karabasov2} have the advantage of being flexible,
with a blending function allowing easy application in complex geometries.

Arguably, the most physically-meaningful choice
for a coupling scheme is one that is built on NEMD theory,
designed to obey the variational forms of the equations of motion.
These aim to control the average properties of a group of molecules in a way that 
minimises the departure from the original unconstrained dynamics.
%
%
There is a long history of developing constraints 
in the NEMD literature.\citep{Hoover, Evans_Morris}
These rely on the variational forms of the equations 
of motion, such as  the principle of least action,
\begin{align}
\delta \textit{A} = \delta \int\limits_{t_1}^{t_2} \left[  \mathcal{L}  + \lambda C \right] dt = 0, 
\label{Constrained_Action}
\end{align} 
where $\mathcal{L}$ is the Lagrangian, 
$C$ is some constraint applied to the system,
and $\lambda$ a Lagrangian multiplier
derived to enforce the desired constraint.
Constraints applied in this manner allow for the dynamics 
of the system to evolve in a physically correct manner 
(minimising the action) while simultaneously satisfying a 
prescribed constraint condition. 
This is of vital importance when the molecular system 
must evolve in a consistent manner with the continuum system.
The constraint $C$ can be either holonomic (a function of position and time only $C(\boldsymbol{r},t) = 0$), 
or non-holonomic (a function of position, velocity, and time, $C(\boldsymbol{r},\dot{\boldsymbol{r}}, t) = 0$).
It is known that \eq{Constrained_Action}, when used in the Euler--Lagrange form,
 does not give the correct equations of motion for non-holonomic constraints,\citep{Saletan_Cromer}
 although some controversy exists.\footnote{\citet{Goldstein} 3rd edition in the errata at \textit{http://astro.physics.sc.edu/Goldstein/} acknowledges several errors and suggests the reference by \citet{Flannery_04}}

This controversy is important as the original 
work of \citet{OConnell_Thompson}, 
used an Euler--Lagrange equation with 
a constraint on the momentum of the MD system, 
see \citet{OConnell_Thesis} for full details.
The constraint of \citet{OConnell_Thompson} drives the MD system until 
it agrees with the continuum.
\begin{align}
 	\COCT (\dot{\boldsymbol{r}}, t)  = M_I \boldsymbol{u}_{I}(t) - \displaystyle\sum_{n=1}^{N_I(t)
 	} m_n \dot{\boldsymbol{r}}_{n}  = 0,
\label{OCTconstraint}
\end{align}
where $M_I \boldsymbol{u}_{I}$ is the continuum momentum
in cell $I$ that overlaps the cell in the MD region 
(refer to Fig.~\ref{fgr:DD_schematic},
where this constrained region is labelled  
`CFD $\to$ MD Boundary condition').
This apparently non-holonomic constraint is, 
in fact, semi-holonomic and can be integrated 
to give a holonomic constraint.\citep{Smith_Thesis}
This semi-holonomic property means applying the constraint
using 
the Euler--Lagrange equation\citep{Goldstein} 
results in the following equations,
\begin{subequations}
\begin{eqnarray}
	\boldsymbol{\dot{r}_{ i}} &=& \frac{\boldsymbol{{p}_{ i}}}{m_i} + \xi \left[ \frac{M_I}{m N_I} \boldsymbol{u_{ I}} - \frac{1}{N_I} \displaystyle\sum_{n=1}^{N_I}\frac{\boldsymbol{{p}_{ n}}}{m} \right] 	\label{OCTConstrnt} \\
	\boldsymbol{\dot{{p}}_{ i}} &=& - \frac{\partial \phi}{\partial \boldsymbol{r_{ i }}} = \boldsymbol{F_{ i}},
	\label{OCTConstrnt2}
\end{eqnarray}
\end{subequations}
written in Hamilton form.
\citet{OConnell_Thompson} introduce a tuning or relaxation coefficient 
$\xi$ to allow the strength of constraint to be reduced.
The term multiplied by $\xi$ is proportional to the 
momentum difference between
the molecular and continuum systems, 
a proportional control in the language 
of control theory. 
In the review of \citet{bian2020domain} 
the work of \citet{OConnell_Thompson} 
is described as `relaxation dynamics', 
and in \citet{Delgado_Buscalioni_2012} as a
Langevin equation.
However, it is important to note the derivation
in the thesis of \citet{OConnell_Thesis} 
is rigorously derived from the principle of least action,
with no stochastic terms, 
and is mathematically and physically identical 
to the form given in the work of \citet{Nie_et_al} 
as shown later in this section.

Despite its rigorous derivation from minimisation principles, 
the constraint of \citet{OConnell_Thompson} 
is missing the localisation in space implied by 
a sum over $N_I$ molecules,
which we include through the $\vartheta$ function 
introduced in Sec.~\ref{sec:MD_ave}.
This subtle difference has two important implications, 
1) the constraint is not semi-holonomic with 
localisation (they depend on position) so it is 
no longer clear if the principle of least action is applicable 
and 2) the explicit localisation results in surface flux 
terms missing in previous derivations.
This localisation can be included by rewriting the 
constraint of \eq{OCTconstraint} in terms of $\vartheta_i$ as follows,
\begin{align}
\CCV(\boldsymbol{r},\boldsymbol{\dot{r}},t)  = \displaystyle\sum_{n=1}^{N} m_n \boldsymbol{\dot{r}}_n \vartheta_n -  \int_V \rho \boldsymbol{u}(t) dV =  0,
\label{Constraint_on_CV_velocity}
\end{align}
Note the continuum is explicitly written in control volume or FV form, 
acknowledging the overlap between continuum and molecular 
must be over a finite volume in space and not a 
differential point.\citep{Smith_et_al, Smith_et_al2015}
The constraint derived from the principle of least action
with explicit localisation is then of the form,
\begin{subequations}
\begin{eqnarray}
	\boldsymbol{\dot{r}_{ i}} &=& \frac{\boldsymbol{{p}_{ i}}}{m_i} + \frac{\vartheta_i}{M_I}  \left[  \displaystyle\sum_{n=1}^{N} \boldsymbol{{p}_{ n}} \vartheta_n -  \int_V \rho \boldsymbol{u} dV  \right]
\label{CVOCTConstrnt} \\
	\boldsymbol{\dot{{p}}_{ i}} &=& \boldsymbol{F_{ i}} + \frac{  m_{i^{\prime}} }{M_I}   \left[  \displaystyle\sum_{n=1}^{N} \boldsymbol{{p}_{ n}} \vartheta_n -  \int_V \rho \boldsymbol{u} dV  \right],
	\label{CVOCTConstrnt2}
\end{eqnarray}
\end{subequations}
where \eq{CVOCTConstrnt2} has a flux term $m_{i^{\prime}}  = \displaystyle\sum_{\alpha=1}^{N_{surf}} m_i \delta(\alpha -x_i) \Lambda_{yi}\Lambda_{zi}  $ 
which is only non-zero when a molecule is crossing 
one of the volume surfaces.
Comparing \eq{CVOCTConstrnt2} to 
the \citet{OConnell_Thompson} equation \eq{OCTConstrnt2}
we see the flux term was missing in previous work 
due to the lack of explicit localisation.
It is likely this omission has not been noticed 
because the proportional control force in \eq{CVOCTConstrnt} removes
any difference between molecular and continuum momenta,
so any error from this missing flux term is corrected at each step. 
However, this missing term becomes essential when we consider 
the commonly used reformulation of 
\citet{OConnell_Thompson} presented in the paper
of \citet{Nie_et_al} In its original form \citet{Nie_et_al} 
obtained this by differentiating 
\eq{OCTConstrnt} and combining with 
equation \eq{OCTConstrnt2} to give a single 
equation in the form,
\begin{align}
m_i \ddot{\boldsymbol{r}}_i = \boldsymbol{F}_i+ \boldsymbol{F}_i^C
\end{align}
where the constraint force is,
\begin{align}
\NCER & = - \frac{1}{N_I} \displaystyle\sum_{n=1}^{N_I}{\boldsymbol{F}_n} + \frac{D\boldsymbol{u}_{I}}{Dt} .
\label{NCRConstntmain}
\\
 & \approx - \frac{1}{N_I} \displaystyle\sum_{n=1}^{N_I}{\boldsymbol{F}_n} + \frac{1}{\Delta t_{MD}} \left[\boldsymbol{u}_I(t+\Delta t_{MD})  -  \frac{1}{N_I} \displaystyle\sum_{n=1}^{N_I} \boldsymbol{\dot{r}}_n(t)  \right].
\label{NCER_prop_constraint}
\end{align}
In combining the equations of \citet{OConnell_Thompson}, 
to get the constraint of \eq{NCRConstntmain}, 
this changes the 
form of constraint to a differential control algorithm.
Differential control aims to ensure the time evolution
of both systems is the same. 
Such constraints typically perform poorly in MD systems,
a well-known problem in the NEMD literature highlighted
by the drift in Gaussian thermostats.\citep{Evans_Morris}
To overcome the limitations of using a differential 
constraint, \citet{Nie_et_al} discretised the time 
derivative in a way that applies a further constraint 
proportional to the velocity in 
both systems, to get \eq{NCER_prop_constraint}. 
This is justified by the requirement that the velocity of the cell at time $t$ 
should tend to the velocity of the 
continuum at time $t + \Delta t_{_{MD}}$, that is, 
\begin{align}
\frac{D\boldsymbol{u}_I}{Dt} \approx \frac{\boldsymbol{u}_I(t+\Delta t_{MD}) -  \boldsymbol{u}_I(t)   }{\Delta t_{MD}} \approx \frac{1}{\Delta t_{MD}} \left[\boldsymbol{u}_I(t+\Delta t_{MD})  -  \frac{1}{N_I} \displaystyle\sum_{n=1}^{N_I} \boldsymbol{\dot{r}}_n(t)  \right].
\label{NCRconstrntDuDT}
\end{align}
The form of \eq{NCER_prop_constraint} can also be obtained 
directly from
a leapfrog discretisation of Eqs.~(\ref{OCTConstrnt}) 
and (\ref{OCTConstrnt2}), 
as shown in the appendix, 
which emphasise the similarity 
between \citet{OConnell_Thompson} and \citet{Nie_et_al}
However, the special discretisation of
\eq{NCRconstrntDuDT} is actually introducing a new 
proportional control, which means the equation is no
longer the form that would be derived from the
principle of least action.
The proportional term then ensures the systems agree
is the same as the main part of the velocity 
controllers of,\citet{Borg_et_al}
\begin{align}
\BORG = K_p \frac{m_i  }{\Delta t} \left[ \boldsymbol{u}(t+\Delta t) - \frac{1}{N_I} \displaystyle\sum_{n=1}^{N_I} \boldsymbol{\dot{r}}_n(t)    \right] ,
\label{Borg}
\end{align}
while the sum of forcing term $\frac{1}{m N_I} \sum_{i=1}^{N_I}{\boldsymbol{F}_i} $ 
is not essential to the functioning of
\eq{NCER_prop_constraint}.
Despite this, later work by \citet{Yen_et_al} proposed 
that the sum of the force terms be averaged 
in \eq{NCRConstntmain} over $M$ iterations to address 
concerns with signal to noise ratios, applied together 
with the time averaged MD velocity 
instead of the instantaneous values 
in \eq{NCRconstrntDuDT},
\begin{align}
\frac{1}{m N_I} \displaystyle\sum_{i=1}^{N_I}{\boldsymbol{F}_i} \approx \bigg\langle \frac{1}{m N_I} \displaystyle\sum_{i=1}^{N_I}{\boldsymbol{F}_i} \bigg\rangle; \;\;\; \frac{1}{N_I} \displaystyle\sum_{i=1}^{N_I} \boldsymbol{\dot{r}}_i(t) \approx  \bigg\langle \frac{1}{N_I} \displaystyle\sum_{i=1}^{N_I} \boldsymbol{\dot{r}}_i(t) \bigg\rangle,  \nonumber
\end{align}
where angular brackets here denote an average over $M \Delta t$.
A further extension of this averaged force model was 
deployed by \citet{Sun_et_al10}, who applied the same force
on all molecules in the overlap region. 
This could potentially have caused problems 
if the continuum profile varied sufficiently rapidly 
in the overlap region as this behaviour would not be 
captured. Borrowing the Quadratic Upstream Interpolation
for Convective Kinetics (QUICK \citep{Hirsch}) scheme
from the continuum literature, 
the force applied was varied linearly across the overlap
region to provide the required velocity profile.
Similarly the temperature was controlled using 
a series of Langevin thermostats with set points 
based on the QUICK scheme.\citep{Sun_et_al10}

In a similar vein, \citet{Wang_He} re-introduced the 
scaling parameter $\xi(t)$ of \citet{OConnell_Thompson} 
to the formula of \eq{NCER_prop_constraint}. 
The $\xi$ parameter was derived as a function of time by
rearranging the constrained equation of motion and the 
constraint was applied gradually over many MD time steps.
Superior performance for noisy simulation
is reported by \citet{Yen_et_al,Sun_et_al10}, 
and \citet{Wang_He} when using averaged or scaled form of
the \citet{Nie_et_al} constraint.
However, these changes represent 
a further departure from the equation obtained 
from the minimisation principles.

To derive a truly localised constraint from minimisation principles, 
we use Gauss' principle of least constraint 
as the constraint of \eq{Constraint_on_CV_velocity} is non-holonomic, 
\begin{align}
\frac{\partial}{\partial \ddot{\boldsymbol{r}}_j} \left[ \frac{1}{2} \displaystyle\sum_{i=1}^{N} m_i \left( \ddot{\boldsymbol{r}}_i - \frac{\boldsymbol{F}_i}{m_i} \right)^2 -  \lambda C \right]=0.
\label{GPLC}
\end{align}
This is because Gauss' Principle, as stated in 
\citet{Flannery_11} p23, is 
\textit{\textit{`a true minimisation principle,}} [...] 
\textit{\textit{with the additional and powerful advantage that
it can be applied to general non-holonomic constraints'}}. 
Equation~(\ref{GPLC}) minimises the local difference between 
forces and acceleration at every time, with any form of 
constraint applied every timestep. The price for this 
generality is the loss of energy conservation ensured
by Hamilton's principle with holonomic constraints.

Applying Gauss' principle to the explicitly localised 
constraint \eq{Constraint_on_CV_velocity} to derive the 
constraint force,\citep{Smith_et_al2015}
\begin{align}
\FCV =& - \frac{m_i \vartheta_i}{M_I} \left[ \frac{d}{dt}\displaystyle\sum_{n=1}^N m_n \dot{\textbf{r}}_n \vartheta_n - \frac{d}{dt}\int_V \rho \boldsymbol{u} dV \right] \label{GLC_EOM} \\
= &- \frac{m_i \vartheta_i}{M_I} \left[   \displaystyle\sum_{\alpha=1}^{Nsurf} \left( \displaystyle\sum_{n^{\prime}}^{N_{\alpha}}  m_{n^{\prime}} \dot{\textbf{r}}_{n^{\prime}} +  \displaystyle\sum_{{nm}^{\prime}}^{N_{\alpha}}  \boldsymbol{f}_{nm^{\prime}}  \right) - \frac{d}{dt}\int_V \rho \boldsymbol{u} dV \right] 
\label{GLC_Stress_full} \\
= &  - \frac{m_i \vartheta_i}{M_I}  \;\; \displaystyle\sum_{\alpha=1}^{Nsurf} \;\; \displaystyle\sum_{n^{\prime}}^{N_{\alpha}}  m_{n^{\prime}} \dot{\textbf{r}}_{n^{\prime}} + \NCER 
\label{GLC_Stress}
\end{align}
The first line, \eq{GLC_EOM}, simply states the time evolution of the molecular 
volume must be subtracted and replaced by the time evolution
of its overlapping CFD counterpart. 
The momentum equation (\eq{mom_eqn})
is then used to obtain \eq{GLC_Stress_full} in terms of surface fluxes.
The final equality gives \eq{GLC_Stress}, to compare to the 
\citet{Nie_et_al} constraint force.
This is obtained by noticing two things,
the first is the sum of $N_I$ forces $\boldsymbol{F}_n$ 
is the same as the sum of forces over the control volume surface
$\sum_{\alpha=1}^{Nsurf} \sum_{{nm}^{\prime}}^{N_{\alpha}}  \boldsymbol{f}_{nm^{\prime}}  = \sum_{n=1}^{N_I}{\boldsymbol{F}_n} $. 
This is because all internal forces between molecules 
inside a control volume are equal
and opposite, so only surface fluxes
are non-zero after the summation over $N_I$. 
The second is that the time evolution of a control volume is a more precise
notation for the substantial derivative. 
As the continuum cell \textit{must} overlap a finite molecular 
volume $D\boldsymbol{u}/Dt$ must apply to a control volume so
$\frac{d}{dt}\int_V \rho \boldsymbol{u} dV \equiv \frac{D\boldsymbol{u}_I}{Dt}$. 
Therefore, we see an additional surface flux term $\sum_{\alpha=1}^{Nsurf} \sum_{n^{\prime}}^{N_{\alpha}} 
 m_{n^{\prime}} \dot{\textbf{r}}_{n^{\prime}}$ when compared to the 
force of \citet{Nie_et_al}


\begin{figure}[bt!]
\centering
  \includegraphics[width=\columnwidth]{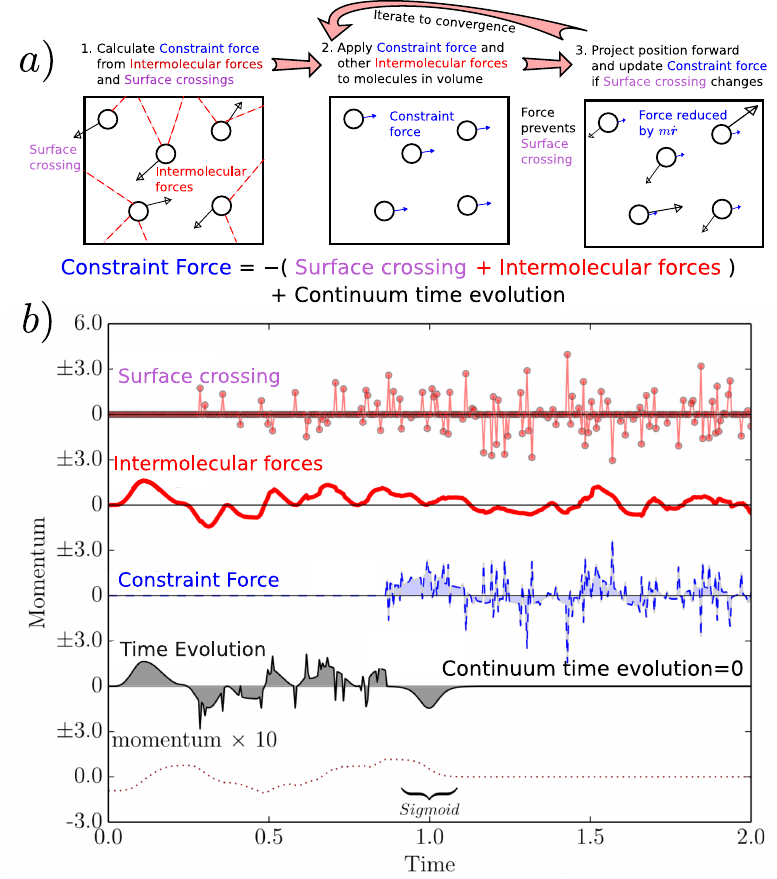}
  \caption{The iterative process required for an exact 
  constraint derived from the principle of least action 
  or Gauss' principle is shown schematically in $a)$,
  while the time evolving results for a control volume 
  in an MD simulation are shown in $b)$ 
  switched on at about time 1.0 to ensure the time evolution
  of the control volume is equal to zero after a signum
  function used to set average momentum to zero. 
  The momentum due to surface crossings is shown on 
  the top line, intermolecular forces crossing the surface 
  on the next line down and the required constraint force in the middle,
  (\textit{i.e.} a force exactly equal to crossings 
  and forces cancels out any momentum change). 
  Details of the complete molecular setup are 
  given in \citet{smith2015localized}}
  \label{fgr:iterate}
\end{figure}

We come to perhaps the most important result of this 
perspective: without this additional flux term, the differential constraint 
of \eq{GLC_Stress} will not work.
The \citet{Nie_et_al} constraint and its derivatives are 
widely used,\citep{Wang_He, Sun_et_al10, Sun_et_al12,  LAMMPS_OPENFOAM, Wu_et_al14, KAMALI_et_al12, yousefi2022}
so this error is significant to a wide range of coupling applications.
It is worth noting, an identical form to \eq{GLC_Stress} is obtained by directly combining 
Eqs.~(\ref{CVOCTConstrnt}) and (\ref{CVOCTConstrnt2}), see \citet{Smith_et_al2015} for details.
Explicit localisation is therefore vital to obtaining the correct constraint.
This also changes the nature of the constraint, the CFD and MD systems must agree as time evolves.
As a result, the constraint becomes iterative in order to ensure the applied force gives the correct momentum at the next timestep.

\begin{figure}[bt!]
\centering
  \includegraphics[height=4.5cm]{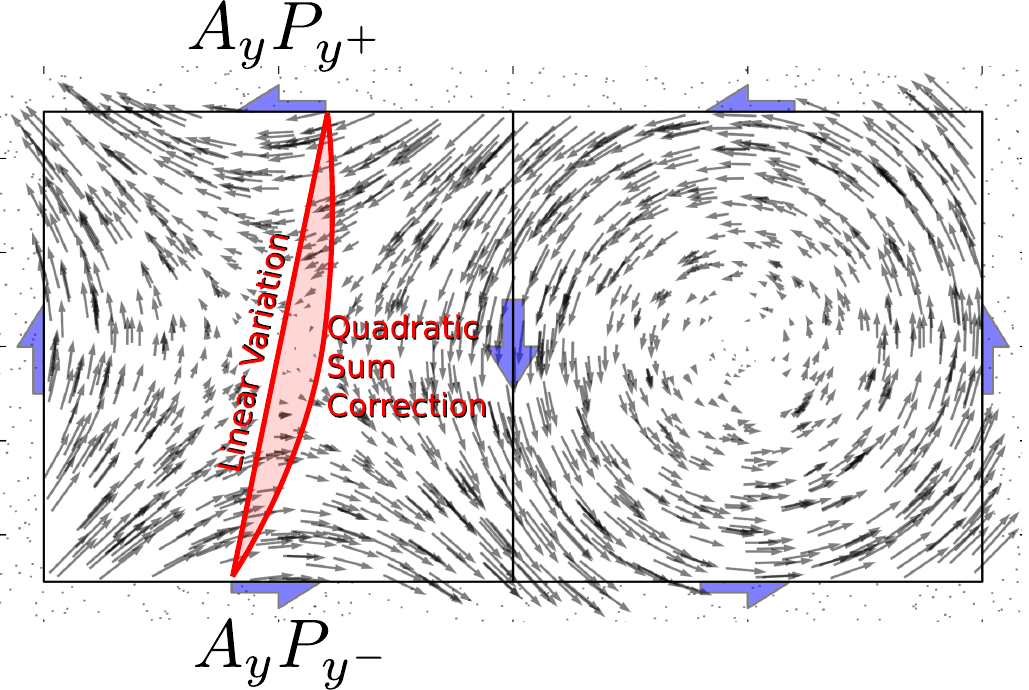}
  \caption{A weighting function constructed to give dilatational flow in the left control volume and rotational flow on the right by choice of surface pressures, with magnitudes of $\Csurf{\!}_{\alpha}$ shown as blue arrows, where each arrow is the velocity vector at the  location of a molecule. The red line shows the linear variation of force in the y direction as discussed in the text while the parabolic addition is chosen to ensure the sum of forces on all molecules obeys the required total momentum change, in this case, the total momentum change in both volumes is zero despite these spatial variations.}
  \label{fgr:weights}
\end{figure}

Implementation of \eq{GLC_Stress} 
requires iteration to ensure fluxes are accounted 
for during the application of the constraint. 
The fluxes are the surface crossings 
(molecules carrying momentum into and out of a volume) 
which are added to the intermolecular forces 
(forces acting over the volume surface).
To understand this, consider the process shown
graphically in Fig.~\ref{fgr:iterate}$a$ for the case 
where the momentum in the CFD system is constant, 
so the continuum time evolution is zero
\ie $\frac{d}{dt}\int_V \rho \boldsymbol{u} dV=0$. 
First, an initial guess for the evolution of molecules over the next timestep
is determined from just the intermolecular forces.
The molecules are projected forward and the fluxes 
measured over the control volume surface.
The constraint force is obtained by summing both 
these fluxes and surface forces according to \eq{GLC_Stress}.
This constraint force is then applied and the
projected evolution of the molecules recalculated.
If any molecules that previously left or entered 
a volume no longer do, the fluxes must be updated.
This in turn changes the constraint force. 
Hence, a new force must be calculated and this process 
iterated until a constraint force 
is consistent with the surface crossings it causes.
Usually, the process takes a few iterations, even with a 3D 
grid of adjacent volumes each iterating 
their own constraint force and fluxes.
The resulting constraint force is shown in 
Fig.~\ref{fgr:iterate}$b$, 
where this force is constructed each timestep to ensure
all momentum change due to fluxes over the volume surface are cancelled out. 
The constraint is turned on just before time $t=1.0$ (simulation units) using a sigmoid
function to guide the constrained volume to a constant momentum value of $\int_V \rho \boldsymbol{u} dV = 0$.
After the constraint is applied, the momentum stays exactly at zero, the 
time derivative of momentum in the volume is kept at zero by the differential constraint
adding a force equal to flux and surface force each time.
Note that intermolecular forces are not changed and each molecule has its own momentum,
the constraint only acts to ensure the total for the volume is tweaked to compensate for any surface
fluxes or forces that could change the momentum inside.
 
Part of the reason the \citet{Nie_et_al} constraint is widely 
used is its simplicity and robustness;
at its core it is just an applied force proportional to velocity difference.
The true form from the variational principles 
should be a differential constraint,
which requires tracking of all surface fluxes and iteration.
Applying such a constraint is more complex, but it is essential to 
apply a minimal constraint consistent with the underlying physics.

\subsubsection{Flux Coupling}
The importance of including fluxes 
naturally brings us to flux coupling.
The first paper on flux coupling
by \citet{Flekkoy_et_al} involved introducing a flux of
molecules at a rate of $\dot{N}=dN /dt$ into the 
molecular domain to ensure mass conservation. 
With molecular actively introduced, 
the flux is therefore easily controlled by choosing 
the momentum of the molecules inserted, $ m\dot{N}\langle \dot{\boldsymbol{r}}' \rangle$.
The velocities $\dot{\boldsymbol{r}}'$ are chosen
randomly from a Maxwell Boltzmann distribution with mean 
value based on the continuum velocity $\boldsymbol{u}$.

Momentum conservation is ensured by
constructing an applied force $\FLEK$ to add up to the 
pressure minus this momentum flux, 
so the resulting constraint force is then of the form,
\begin{align}
\FLEK = \frac{g(y_i)}{\displaystyle\sum_{i} g(y_i)}  \left[ A \boldsymbol{P} \cdot \boldsymbol{n} - \displaystyle\sum_{n^{\prime}}m_{n^{\prime}} \dot{\textbf{r}}_{n^{\prime}}     \right] .
\label{Flekkoy_constraint}
\end{align}
Here $g(\boldsymbol{y}_i)$ is a weighting with 
an arbitrary function form based on distance from 
the top of the domain. 
The flux term appearing in \eq{Flekkoy_constraint} 
from \citet{Flekkoy_et_al} appears to be equal to the
molecular insertion, $\sum_{n^{\prime}}m_{n^{\prime}} \dot{\textbf{r}}_{n^{\prime}}  = m\dot{N}\langle \dot{\boldsymbol{r}}' \rangle$,
although more generally this could be any molecule 
entering the region where the constraint is applied.

\citet{Delgado-Buscalioni_Coveney_03} extended this model
by introducing an energy-based flux for the purpose of 
simulating unsteady flow.
The weighting function 
was set to unity, $g(\boldsymbol{y}_i)=1$,
 so that the applied force was the same for all 
molecules to allow easier control on the energy added as external work done becomes,
$\sum \boldsymbol{F}_i^{C}{\boldsymbol{v}}_i = \boldsymbol{F}^{C} \cdot \boldsymbol{u}$.
The correct energy flux is ensured by inserting molecules 
with the appropriate kinetic energy from the Maxwell 
Boltzmann distribution and at a location that matches 
the required potential energy. 
Finding the required energy for molecular insertions is a complicated aspect of coupling, 
and is discussed in Section~\ref{sec:insertion}.
The conduction (as a temperature gradient) is matched 
to the continuum using a series of thermostats.

Flekk\o{}y, Delgado-Buscalioni and Coveney \citep{Flekkoy_DB_Coveney} later reformulated this flux 
constraint to include a fluctuating part to control 
the energy addition directly,
\begin{align}
\FDBC = \FLEK + \boldsymbol{F}^{C\prime}.
\label{FDBC_force_stress}
\end{align} 
The constant part, $\FLEK $, is identical to 
\eq{Flekkoy_constraint} with weighting function equal to unity 
while the fluctuating term, $\boldsymbol{F}^{C\prime}$, adds no net momentum,
instead providing only energy,
\begin{align}
\boldsymbol{F}^{C\prime} = \frac{\boldsymbol{v_i}}{\sum_{i=1}^{N_I} \boldsymbol{v}_i^{2}} \left( \mathcal{E} A - \displaystyle\sum_{i^\prime} \epsilon_{i^{\prime}}  - \FLEK \cdot \boldsymbol{u}  \right). \
\label{Energy_FDBC}
\end{align} 
The magnitude of fluctuating force applied to each particle
is based on the particles thermal energy, 
\textit{i.e.} minus streaming term
$\boldsymbol{v}_i = \boldsymbol{\dot{r}}_i - \boldsymbol{u}$.
The proposed force is said to be derived with the aim of ensuring a
reversible force, one that adds no energy and ensures 
the probability distribution
$\textit{f}^{eq} = exp (-k_B \mathcal{H}/ T) / Z$ is 
preserved at every time step, where $Z$ is the partition
function, $\mathcal{H}$ the Hamiltonian, and 
$k_B$ is Boltzmann's constant.
Later work by the same group, \citep{DeFabritiis_et_al_07,Delgado-Buscalioni_DeFabritiis}
replaced the continuum solver with the equations of 
fluctuating hydrodynamics.\citep{Landau_lifshitz_fluidmech}
These stochastic equations add an extra noise term to retain the 
small scale fluctuations in the continuum solver. 
The noise term is generated 
using a Wiener process and was tuned to satisfy the 
fluctuation--dissipation theorem. 
This allows molecular fluctuations to be preserved 
in the continuum part of the solver. 
These flux coupling developments are summarised in a 
review by \citet{Delgado_Buscalioni_2012}.



We can show the link between state and flux coupling,
and, at the same time, show a direct derivation of flux coupling 
starting from Gauss' principle of least constraint. 
Recognising the continuum time evolution can be written
in terms of surface fluxes over $N_{surf}$ surfaces $\frac{d}{dt}\int_V \rho \boldsymbol{u} dV = \sum_{\alpha=1}^{Nsurf}  \int_{A_{\alpha}} \boldsymbol{P} \cdot d \textbf{A}_{\alpha}$,  
we obtain from \eq{GLC_Stress_full},
\begin{align}
\FCV =  \frac{m_i \vartheta_i}{M_I} \displaystyle\sum_{\alpha=1}^{Nsurf}  \underbrace{\left[  \int_{\textbf{A}_{\alpha}} \boldsymbol{P} \cdot d \textbf{A}_{\alpha} - \displaystyle\sum_{n^{\prime}}^{N_{\alpha}}  m_{n^{\prime}} \dot{\textbf{r}}_{n^{\prime}} -  \displaystyle\sum_{{nm}^{\prime}}^{N_{\alpha}}  \boldsymbol{f}_{nm^{\prime}}   \right] }_{\Csurf{\!}_{\alpha}}. 
\label{GLC_flux}
\end{align}
We introduce the notation $\Csurf{\!}_{\alpha}$ to highlight 
this is in the form of a constraint minimising the difference 
between the continuum and molecular pressures on a surface $\alpha$.
It is instructive to compare to the flux form
of \citet{Flekkoy_et_al} as shown
in \eq{Flekkoy_constraint}, 
chosen with the arbitrary weighting function as $g(\boldsymbol{r}) = m_i \vartheta_i$
and noting $M_I = \sum_{i=1}^N m_i \vartheta_i$. 
As the constraint of \eq{GLC_flux} is derived 
for a control volume, it requires the sum of fluxes over all the surfaces of that
volume ($6$ for a cuboid) in order to constrain
the momentum inside that volume.
To understand this in terms of the momentum constraint of \eq{Flekkoy_constraint}, 
we consider the typical geometry of application 
used by \citet{Flekkoy_et_al} \cite{Flekkoy_DB_Coveney}
As flux constraints are applied to a buffer of molecules terminating 
with an open boundary to a vacuum at the domain top, 
no intermolecular forces would exist on the top surface $y^+$ 
and insertion is used to ensure the required momentum agrees between continuum and molecular. 
As a result, the momentum agreement on the top surface $y^+$ would be automatically ensured so does not appear in the constraint equations.
Assuming periodic boundaries in the other directions,
fluxes on connected faces $x^+$ to $x^-$ and $z^+$ to $z^-$ would cancel.
As a result, only fluxes on the bottom surface need to be considered 
in the applied force to ensure momentum control of the volume,
\begin{align}
\FCV =    \FLEK  -  \frac{m_i \vartheta_i}{M_I}\displaystyle\sum_{{nm}^{\prime}}^{N_{y^-}}  \boldsymbol{f}_{nm^{\prime}}.
\label{GLC_Flekkoy}
\end{align}
For this particular geometry, the form of flux constraint 
from \citet{Flekkoy_et_al} can be considered to be identical 
except for an additional 
intermolecular force term $\boldsymbol{f}_{nm^{\prime}}$.
It is natural to ask why this additional term is not essential for
flux based coupling to work successfully, as shown by various publications.
\citep{Flekkoy_et_al, Flekkoy_DB_Coveney, Werder_et_al, DeFabritiis_et_al_06, DeFabritiis_et_al_07, Delgado_Buscalioni_2012,  Papez_Praprotnik_2022}
It is possible the impact of this missing intermolecular
force term requires a correction, as used in \citet{Delgado_Buscalioni_2012} 
applied to the whole volume in order to
ensure conservation between CFD and MD.
It is also possible this term, which depends on molecular configuration, 
is zero on average even for cases of strong flows.
Most likely is that the form of constraint in $\FLEK$ has a feedback structure,
so any difference between molecular momentum flux and continuum pressure 
is applied as a force to drive the flow (ensuring they agree).
However, 
if an exact control on the momentum is required, or an application needs constraint in a region
which is not the entire top of the simulation, then every single force and
flux must be accounted for with iteration as described in 
Fig.~\ref{fgr:iterate} applied.

We have not 
considered the energy control introduced in \eq{Energy_FDBC}.
As the force in \eq{GLC_flux} is derived from 
Gauss' principle with a non-holonomic constraint, 
it necessarily adds energy to the system.
In the limit of zero volume size, it can be shown
that \eq{GLC_flux} adds the same energy to the system as 
the SLLOD equations of motion, important as 
SLLOD was derived to ensure, among other considerations,
that the correct work is done on the MD system.
More generally, we can ask if a momentum constraint force
should add additional energy to the MD?
Coupling to a CFD solver puts that continuum domain
outside of the MD domain, so a coupling constraint would be
expected to do work on the MD system in order to drive it.
The non-holonomic nature of the constraint supports the conclusion, 
\ie a local control of momentum as required for coupling,
makes energy addition inevitable.
If the continuum problem requires coupling of the energy equation,
then energy control will be needed at the interface.
The added constraint of \eq{Energy_FDBC} used
in \citet{Flekkoy_DB_Coveney} aims to control 
both stress heat and energy flux.
There is no reason this could not be included in the extended control volume approach
discussed here, provided care is taken to ensure momentum control is respected.
This could be built in as an additional constraint on energy added to Gauss' principle.

\begin{figure*}
\centering
  \includegraphics[width=\textwidth]{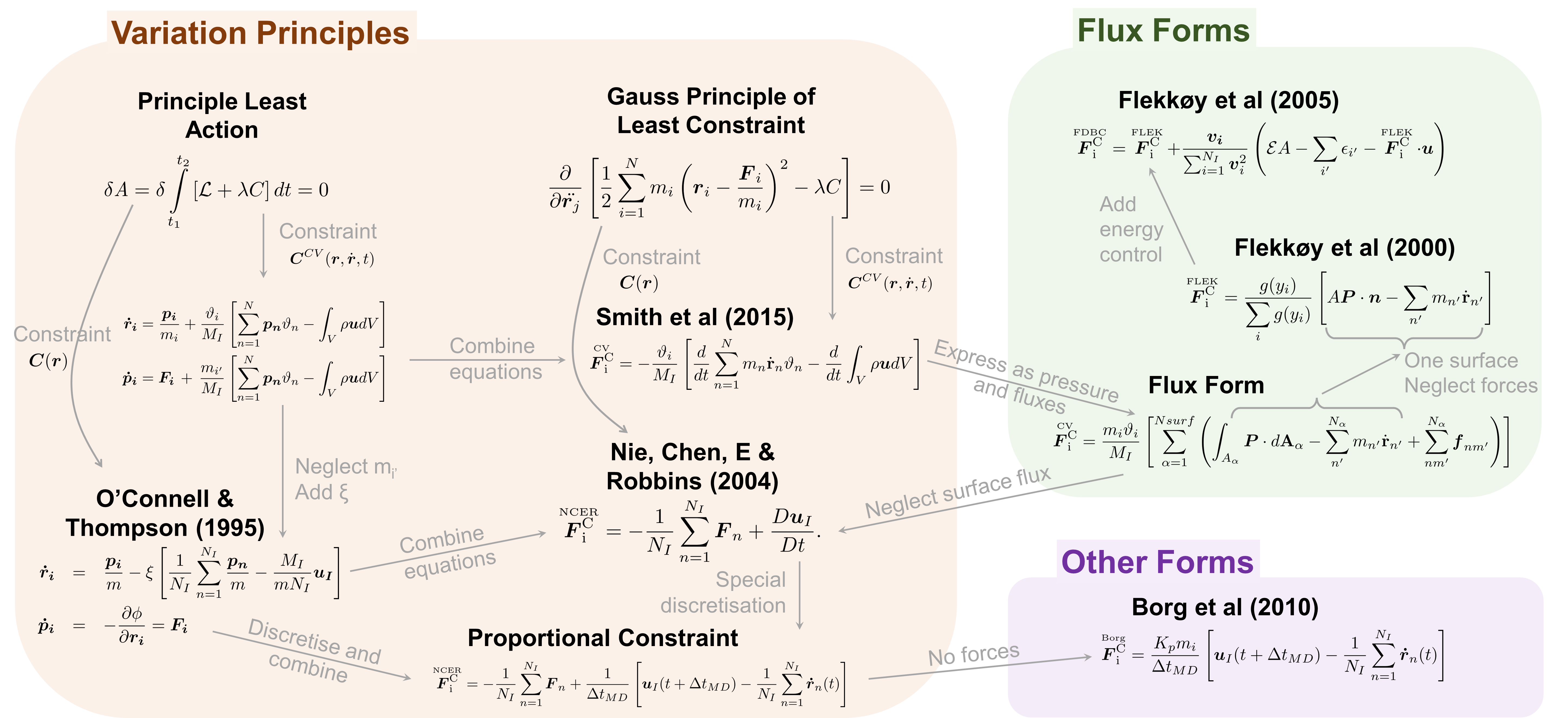}
  \caption{An overview of the mathematical link between the different forms of coupling constraint in the literature. }
  \label{fgr:schematic_constraint}
\end{figure*}
%

This section has shown the sum of fluxes over all surfaces of a volume,
with iteration, is required to enforce the correctly localised momentum constraint derived from Gauss' principle.
For the specific geometry of a single constrained region at the top of the domain,
only a single surface needs to be controlled and this is equivalent to the flux coupling forms 
presented in the literature, as shown in \eq{GLC_Flekkoy}. 
However, the iterative and multi-surface nature of the control volume flux constraint 
allows us to go beyond this single-controlled region and provide exact control 
over all surfaces in a grid of contiguous volumes.
To do this, we use a finite element 
approximation to \eq{GLC_flux} in order to distribute the forces 
with a weighting function $g(\boldsymbol{r})$ allowing us 
to specify the fluxes over each surface independently.
This must be designed to ensure the sum over the volume still satisfies
the momentum constraint condition.\citep{smith2015localized} 
Perhaps the simplest form of weighting function to 
achieve this in a cuboidal volume is 
$g(\boldsymbol{r}_i) = h(\boldsymbol{r}_i)+ \eta f(\boldsymbol{r}_i) $ 
where the vector position denotes the product of the
components in each dimension, \eg 
$h(\boldsymbol{r}_i) = h(x_i) h(y_i) h(z_i) $ and
 $f(\boldsymbol{r}_i) = f(x_i) f(y_i) f(z_i)$.
Using a linearly varying weight between surfaces, \eg in $x$,
$h(x_i) = (\Csurf{\!}_{x^+} - \Csurf{\!}_{x^-} ) \tilde{x}_i + \Csurf{\!}_{x^-}$ 
for $0 < \tilde{x}_i < 1$ or expressed in terms of the commonly used finite 
element shape functions between bottom $x^-$ and top $x^+$ position,
$h(x_i) = \Csurf{\!}_{x^+} N^+(x_i) -  \Csurf{\!}_{x^-} N^-(x_i)$ where 
$N^+(x_i) = [x^+ - x_i] / \Delta x$ and 
$N^-(x_i) = [x_i - x^-] / \Delta x$ with $\Delta x = x^+ - x^-$. \citep{Zienkiewicz}
The added term is constructed to be zero at the surfaces
$f(x_i) = \tilde{x}_i^2 - \tilde{x}_i $ 
or in general coordinates,
$f(x_i) = x_i^2 - x_i (x^+ + x^-) + x^+ x^- $.

The  $\eta$ function is then chosen to ensure that
$\sum_{i=1}^N g(\boldsymbol{r}_i) = 1 $, which requires,
\begin{align}
\eta = \frac{1 - \displaystyle\sum_{n=1}^N h(\boldsymbol{r}_n) \vartheta_n  }{\displaystyle\sum_{n=1}^N f(\boldsymbol{r}_n)  \vartheta_n},
\label{eta_correct}
\end{align}
Notice that the form of $h$ could be changed to any 
functional form, for example a higher order element or 
even the radial distribution forcing used in \citet{Werder_et_al}, 
and \eq{eta_correct} 
would still ensure total weighting sums to unity.
This constraint allows the flux over all surfaces of a control
volume to be controlled, giving complicated flow-fields as shown in Fig~\ref{fgr:weights}. 
An example of using this function to varying stress control in one dimension is shown in the  appendix.

In the most general case, three flux components on six surfaces can be constrained 
allowing $18$ stresses and three momentum values to be enforced on the MD system.
The distribution functions of \eq{GLC_flux} could also be chosen to control other quantities,
for example aiming for a particular mass flux 
(\eg $\Csurf{\!}_{\alpha}=\int_{\textbf{A}_{\alpha}} \rho \boldsymbol{u} \cdot d\textbf{A}_{\alpha} - \sum_{n^{\prime}}^{N_{\alpha}}  m_{n^{\prime}} = 0$).
This could entail controlling the linear variation of pressure so as
to ensure the mass flux matches at the CFD--MD interface.
Seen through this lens, the iterating required to enforce
the constraint shown in Fig.~\ref{fgr:iterate} 
is analogous to the iteration used to enforce
mass continuity in a CFD pressure solver by controlling pressure.
Given the extensive work done on numerics and pressure 
solvers in the CFD community over almost seventy years, 
further work is certainly justified to develop 
such coupling framework further.
Especially for multi-phase, thermal or visco-elastic models,
where the coupling requirements become more complex, distribution of
forces provides a method to control the MD system.
More generally, the presented framework here links the main coupling approaches, 
as summarised in Fig.~\ref{fgr:schematic_constraint}
and provides a potential starting point for a theoretical development to address
more complex coupling requirements.
It also has the potential to solve long-standing problems in embedded style coupling of applying Lees Edwards in 3D \citep{Yamamoto}
by allowing full control of the stress tensor in all directions.

\subsection{Molecular Insertion}
\label{sec:insertion}

\begin{figure}[ht!]
\centering
  \includegraphics[width=\columnwidth]{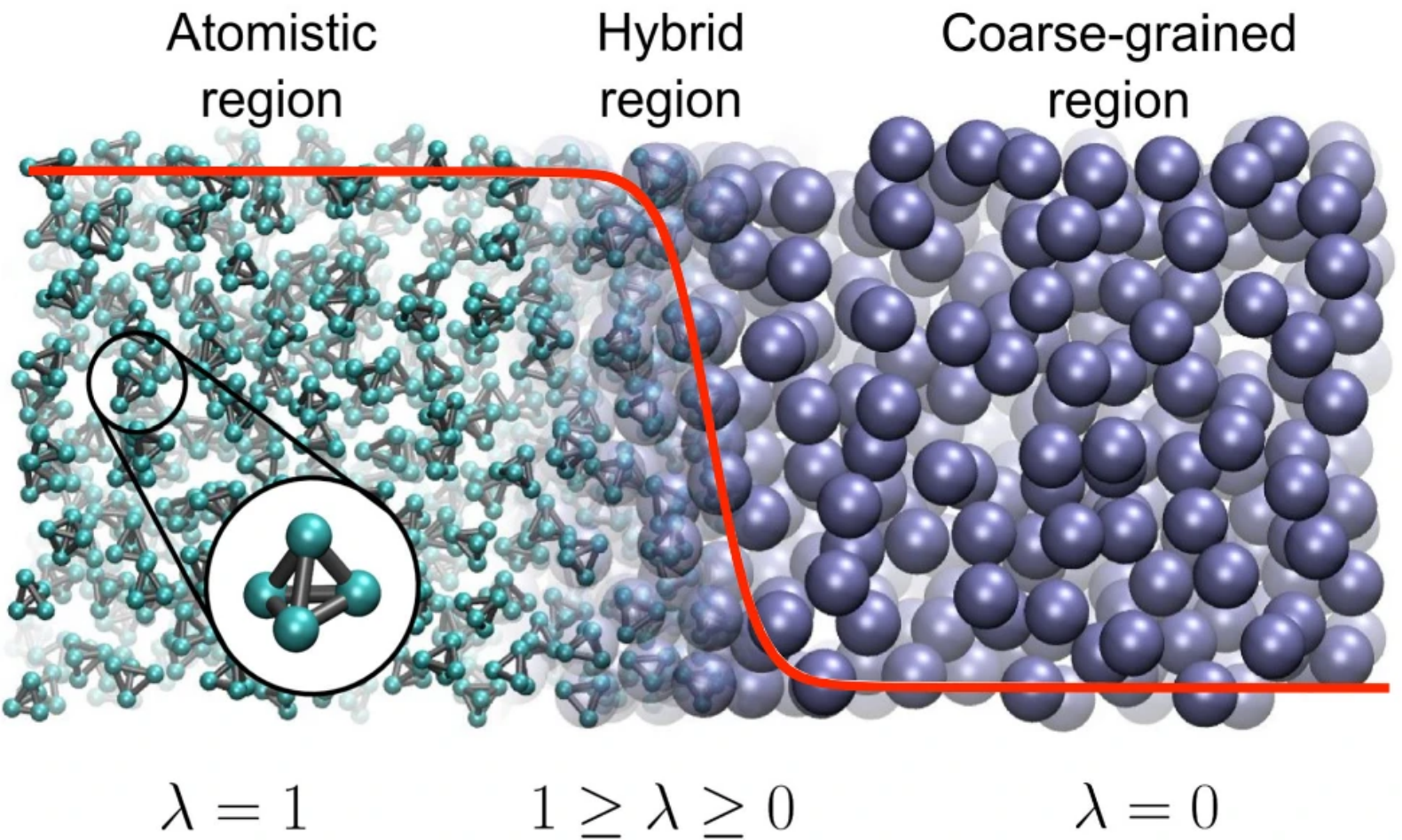}
  \caption{Illustration of an AdResS system setup for
  tetrahedral molecules that can move freely between
  the atomistic and the CG regions through a
  hybrid region as indicated.
  Reprinted figure with permission from Ref.~\cite{Potestio2013}.
Copyright (2013) by the American Physical Society.}
  \label{fig:1}
\end{figure}

Molecular insertion has been the focus of extensive research, 
we briefly outline the main developments here,
and refer readers to \citet{CortesHuerto2021} for a recent review.
The work of \citet{OConnell_Thompson} did not use particle 
insertion as a force is applied to stop molecules escaping, while \citet{Nie_et_al}
used the gap created by this force to make molecular insertion straight forward.
In \citet{Flekkoy_et_al} the method used to insert particles is not stated explicitly,
but later papers based on the same flux 
coupling\citep{Delgado-Buscalioni_Coveney_03} use a steepest decent
algorithm (USHER) to insert atoms
at a location that gives the required potential energy.\citep{Delgado-Buscalioni_Coveney_03_USHER} 
For atoms and even simple molecules, this works well as it is often possible to find locations.
More complex molecules, especially with long-chain are not possible to insert in this manner.
One approach is to gradually increase the additional detail of these complicated molecules, forcing a
region to accommodate them, as presented in the FADE algorithm.\citep{Borg_et_al_FADE}
The most mature method for complex molecular insertion is the adaptive resolution scheme (AdResS).
In this method, one part of the
system is treated at the all-atom level and
another part at the
coarse-grained (CG)
level, thus allowing on the fly exchange
of molecules between
the atomic and CG levels of
description through a hybrid region
(Fig.~\ref{fig:1}).\cite{Praprotnik2005,Praprotnik2008}
%

In the context of fluids, examples of the 
AdResS approach refer to the simulation
of liquid water\cite{Praprotnik2007,Potestio2013},
which serves as a proof-of-concept for further
applications.
In Ref.~\citen{Praprotnik2007}, 
a TIP3P model was
used for the all-atom representation of water
and a respective CG description as well.
We should however underline that the development
of all-atom models for water with CG force-fields 
is still a very challenging aspect. 
For this, intensive research
has led to the development of different models
with each reproducing a certain range
of water properties.
In any case, a CG model can
be obtained by the all-atom model by bottom-up
approaches, for example, by
matching various dynamic and structural
properties (\textit{e.g.} 
using inverse methods \cite{Reith2003}), 
as well as, (thermodynamic) properties,
such as pressure, \textit{etc.} 
between the different levels of descriptions. 
Top-down approaches are also common, as in
the case of MARTINI \cite{Souza2021} 
and SAFT\cite{muller2014force,Lafitte2013} force-fields.
In the context of AdResS, for example,
a dynamic clustering algorithm that 
concurrently couples atomistic and
CG representations
has been applied in the case of a hybrid
SPC/MARTINI model.\cite{Zavadlav2016,Zavadlav2019}
In this case, the existence of a hybrid regime
can act as a glue between popular force fields,
such as the SPC/E\cite{Berendsen1987}
and the MARTINI.\cite{Souza2021}
Finally, the link to the continuum has been demonstrated
in the simulation of molecular liquids via
a triple-scale simulation.\cite{Delgado2008}
In this case the all-atom and CG descriptions
is coupled via the AdResS scheme, while the
CG level is coupled to the continuum model.
Recipes to address the insertion of large molecules
in the hybrid particle--continuum have been proposed, 
while the model seems to describe the hydrodynamics
of the system.\cite{Delgado2008}
The AdResS scheme has been also used with mesoscale models,
such as dissipative particle dynamics (DPD),\cite{Papez2022}
where the exchange of the information between the
domains is based on the open boundary method.
\cite{Flekkoy2005,Buscalioni2015}
While there are various versions of DPD
models, these models use particle descriptions,
which renders the AdResS framework generally
suitable for this type of
coupling.

\begin{figure}[bt!]
\centering
  \includegraphics[width=\columnwidth]{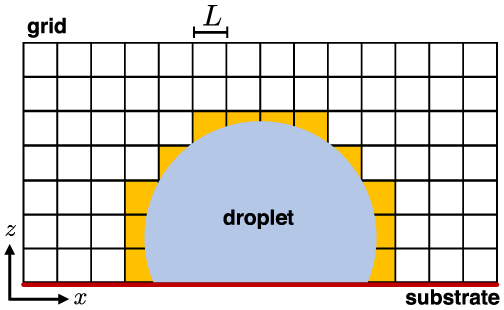}
  \caption{Droplet on a substrate simulated via a molecular-scale MC method. The VOF in the
  cell is used to track the interface of the droplet, where exchange of particles takes place to investigate
  liquid--vapour equilibrium, condensation or evaporation
  phenomena.
From Ref.~\cite{Theodorakis2021}.
}
  \label{fig:4}
\end{figure}

\subsection{Applications}

The early work on coupling typically focused on canonical flows 
such as the Couette and Poiseuille solutions of the Navier--Stokes equations.
The flux coupling of \citet{Flekkoy_et_al} was tested to 
simulate steady state Couette flow using molecular regions
at the top and bottom of the domain and a continuum region
between them, as well as steady state Poiseuille flow
with a molecular region simulating the 
length of the channel (including both walls) in the 
streamwise direction. This flux exchange method was 
extended to include conservation of 
energy for force-driven flow over a flat wall.\citep{Wagner_et_al} 
\citet{Sun_et_al10} apply their model to Poiseuille flow
with energy exchange and later to a wall of equally 
spaced posts.\citep{Sun_et_al12}
The work of \citet{yan_et_al_2021} represents an application
of domain decomposition coupling to explore slip of a 
polymer melt where the near-wall region is modelled 
with MD and the remaining domain by CFD.
They show that as shear rate drops, 
the computational saving increases up to two orders 
of magnitude. 
This use of coupling avoids the need for wall models and they
show the slip measures agrees very well with previous
pure MD simulations.
Applying coupled simulations using complex molecules for
tribology applications has been
discussed in \citet{Edu_Thesis}, while
the consideration of developing slip models from 
coupling was given in \citet{yang2018}
A variety of different materials with varying textures 
linked to varying interaction with fluids is 
considered in \citet{yousefi2022} 
Heat transfer was the focus of recent 
coupling work,\citep{liu_2021}
while a specific application with LAMMPS and OpenFOAM is 
presented in \citet{LAMMPS_OPENFOAM}.

The early work of \citet{Nie_et_al} also explored 
Couette flow and built in a post to induce a flow in the
wall-normal direction. They extended this work later to 
lid-driven cavity flow \citep{Nie_et_alb} to explore the singularity
present where the stationary and moving walls meet.
The \citet{Nie_et_al} model has been applied to a large 
coupled simulation by \citet{Yen_et_al}
They simulated a
large scale Couette flow, 
 an order of magnitude larger than 
that of \citet{Nie_et_al} with a proportionally 
smaller shear rate. 
Start-up Couette flow was also simulated by \citet{KAMALI_et_al12}
with fluid flows for various micro- and
nano-scale geometries studied.
\citet{Delgado-Buscalioni_Coveney_03} simulated an 
oscillating wall (Stokes 2$^{\rm nd}$ problem), which is a rare 
example of an unsteady problem. 
More extensive tests of unsteady coupling modelled include the flow of a shockwave between
domains.\citep{Delgado-Buscalioni_DeFabritiis}
Recently, such application have been extended to 
practical problems such as the transmission of 
ultrasound.\citep{Papez_Praprotnik_2022}
More complex flow past nano-tubes in the form of
cylinders was considered in \citet{Werder_et_al}, although only for creeping flow.
It would be possible to push these simulations into the unsteady 
regimes, which have been shown to be possible using pure MD.\citep{Rapaport}
It is also possible to extend coupled simulation to turbulent flow,\citep{Smith2020} 
where the pure MD case \citep{Smith_2015} can be seen to reproduce turbulence 
at the nanoscale and so a coupled model could allow quick 
optimisation of molecular wall effects on 
turbulent structures.

The blending function approach of Eqs~(\ref{Karabasov})
and (\ref{Karabasov}) lends itself well to complex problems,
such as the diffusion of a biomolecules in water due 
to Couette flow\citep{Hu_et_al_2019} and a PCV2 virus 
capsid in water,\citep{Tarasova_2017} as well an
exploration of the violation of continuum laws in 
atomic force microscopes.\citep{Li_et_al_2023}




In the context of multi-phase flow, 
the moving contact line was considered as early as 
1999,\citep{Hadjiconstantinou_99} 
with later simulations of coupled droplets \citep{Wu_et_al14} 
and droplet impinging on a surface.\citep{Zhou_et_al14}
Another example uses MD simulations to generate 
data which is sent to a phase-field model
based on the Helmholtz energy equation of state
and evaluated by CFD.\cite{Heinen2022}
The volume of fluid (VOF) method appears as a simple and 
robust approach to identify the interface
between two different phases, for example
between a liquid and a gas phase, due
to the large density differences between the
phases (Fig.~\ref{fig:4}). 
A scheme, such as VOF, can be
combined with an off-lattice MC
approach to simulate evaporation and
nucleation phenomena at the molecular 
level.\cite{Theodorakis2021} 
The overall coupling protocol
allows for the exchange of particles
at the interface without the need to
simulate gas molecules far from the
liquid--gas interface and can be used
with any force-field, be it all-atom
or CG, thus allowing the
simulation of a broad range of complex
liquids, for example, nanofluids.\cite{Liu2020}








\section{Conclusions and Future Perspectives}

A great number of coupling possibilities
can be realised between currently available
simulation methods. The list of available methods
(and acronyms) is quite long, for example,
MC, MD, DFT, VOF, FV, FEA, LB, \textit{etc.}
The possibilities are at least as many as the possible 
combinations of these methods and a great amount
of work has been dedicated to linking the
various methodologies in the most computationally
efficient way and as close as possible obeying
the physical laws. Here, we have not
attempted to provide a detailed description 
of these methods, but rather provide a 
perspective on coupling efforts in a very focused area:
MD coupled to continuum methods for 
fluid dynamics, in particular the popular finite volume (FV) method.
Coupling time scales is important here but has been discussed in the
literature.\citep{Delgado-Buscalioni_Coveney_04, Lockerby_et_al13, Delgado-Buscalioni2015}
Instead, the focus is on an area where less progress has been made, developing a theoretical 
framework for domain decomposition coupling, summarised as follow:
Using an explicit localisation function based on the FV form applied to an MD system
results in the form of fluxes on the surface, the MOP pressure, 
which avoids the well-known errors associated with the virial pressure.
This description in terms of surface pressure and fluxes is consistent with the FV 
method used in the CFD and can be shown to be exactly 
conservative in an MD system.
Applying this FV localisation to the derivation 
of a constrained dynamics algorithm results in an new surface flux term,
exposing an error in the central works on coupling 
\citep{OConnell_Thompson,Nie_et_al} and 
consequently all subsequent papers.
The corrected constraint is differential in nature, 
requiring iteration to ensure the time evolution of both
systems match.
This general constrained form can be simplified to different 
well-known expressions from the literature, summarised in Fig \ref{fgr:schematic_constraint}, 
including the derivation of the flux forms from variational principles.
Extending to the a finite element form gives a generalised flux coupling which can be 
applied to every surface of a volume in space, not just the domain top.
Comprehensive control using, for example all 18 surface components of 
pressure, is possible and provides a template for a more general class of coupling methods.
Attempts to overcome this
theoretical barrier through artificial-intelligence approaches are already taking
place \cite{Raissi2019} and these developments should provide a groundwork to
build models on.
These insights are presented in the hope that they will be a stepping stone
for further work and ideas in the development of a rigourous groundwork for coupled simulation.

\section*{Conflicts of interest}
There are no conflicts to declare.

\section*{Acknowledgements}
This research has been supported by the 
National Science Centre, Poland, under
grant No.\ 2019/34/E/ST3/00232.
We gratefully acknowledge Polish high-performance 
computing infrastructure PLGrid (HPC Centers: ACK Cyfronet AGH) 
for providing computer facilities and support 
within computational grant no. PLG/2022/015261.

\section*{Appendix}

\subsection*{Discretising O'Connell and Thompson}

A discretisation of Eqs (\ref{OCTConstrnt}) and (\ref{OCTConstrnt2}) of \citet{OConnell_Thompson} using the leapfrog scheme shows,
\begin{subequations}
\begin{eqnarray}
	\boldsymbol{{r}_{ i}}(t+\Delta t) = \boldsymbol{{r}_{ i}}(t) + \Delta t \bigg( \frac{\boldsymbol{{p}_{ i}} (t+\Delta t/2)}{m_i} \nonumber \\
 \;\;\;\;+ \xi \left[ \frac{M_I(t)}{m N_I(t)} \boldsymbol{u_{ I}}(t+\Delta t) - \frac{1}{N_I(t)} \displaystyle\sum_{n=1}^{N_I(t)}\frac{\boldsymbol{{p}_{ n}}(t+\Delta t/2)}{m} \right] \bigg) 	\nonumber \\
	\boldsymbol{{{p}}_{ i}} (t+\Delta t/2) = \boldsymbol{{{p}}_{ i}} (t-\Delta t/2) + \Delta t \boldsymbol{F_{ i}}(t),
	\label{OCTConstrntDiscretised2}
\end{eqnarray}
\end{subequations}
these can then be combined to give,
\begin{align}
	\frac{\boldsymbol{r}_{i}(t+\Delta t) - \boldsymbol{r}_{i}(t)}{\Delta t}=  \frac{\boldsymbol{p}_{i} (t-\Delta t/2)}{m_i} + \Delta t \boldsymbol{F}_{ i}(t) - \frac{\Delta t }{N_I(t)} \displaystyle\sum_{n=1}^{N_I(t)} \boldsymbol{F}_{n}(t)  \nonumber \\
+ \xi \left[ \frac{M_I(t)}{m N_I(t)} \boldsymbol{u}_{I}(t+\Delta t) - \frac{1}{N_I(t)} \displaystyle\sum_{n=1}^{N_I(t)} \frac{\boldsymbol{p}_{n} (t-\Delta t/2)}{m_n}\right] ,
\end{align}
where for unit mass we have $M_I = m N_I$, setting $\xi=1$, replacing the momentum notation $\boldsymbol{p}_{i}/m_i = \dot{\boldsymbol{r}}_i$ and used the first-order backward Euler finite difference approximation $\dot{\boldsymbol{r}}_i (t-\Delta t/2) / \Delta t = (\boldsymbol{r}_i (t) - \boldsymbol{r}_i (t-\Delta t))/(\Delta t)^2$ and second derivative $\ddot{\boldsymbol{r}}_i = [\boldsymbol{r}_{i}(t+\Delta t) - 2\boldsymbol{r}_{i}(t) + \boldsymbol{r}_{i}(t-\Delta t)]/(\Delta t)^2$ , the form can be seen to be identical to \eq{NCER_prop_constraint} from \citet{Nie_et_al} with the velocity at the half step consistent with the leapfrog scheme,
\begin{align}
\ddot{\boldsymbol{r}}_i 	=  \boldsymbol{F}_{ i}(t) - \frac{1}{N_I(t)} \displaystyle\sum_{n=1}^{N_I(t)} \boldsymbol{F}_{n}(t)  + \frac{1}{\Delta t}\left[ \boldsymbol{u}_{I}(t+\Delta t) - \frac{1}{N_I(t)} \displaystyle\sum_{n=1}^{N_I(t)} \dot{\boldsymbol{r}}_n (t-\Delta t/2)\right].
\end{align}
Note that $N_I$ is itself a function of time and dependent on the molecular position. 

\subsection*{Understanding the Blended Region}

The constraint force of \citet{Markesteijn_et_al_2014} half way across the blending region e.g. for $s=0.5$ is shown here,
\begin{subequations}
\begin{eqnarray}
 \!\!\! \!\!\! \!\!\! \!\!\!	\boldsymbol{\dot{r}_{ i}} & =& \!\!\! \frac{\boldsymbol{{p}_{ i}}}{2m_i}+ \frac{\rho \boldsymbol{u} - [\rho \boldsymbol{u}]^{MD} }{\rho - \rho^{MD}}+ \frac{\alpha }{8 \rho^{MD}} \frac{\partial}{\partial \boldsymbol{r}} \left(\rho - \rho^{MD} \right)  \nonumber \\ 
  \!\!\!\!\!\! \!\!\! \!\!\!\boldsymbol{\dot{{p}}_{ i}} &=&  \frac{\boldsymbol{F}_i}{2} +  \frac{1}{8 \rho^{MD}}\frac{\partial}{\partial \boldsymbol{r}} \cdot   \bigg( \alpha \frac{[\rho \boldsymbol{u}]^{MD}}{\rho^{MD}}\frac{\partial}{\partial \boldsymbol{r}} \left(\rho - \rho^{MD} \right) \dots   \nonumber 
\\ & & \;\;\;\;\;\;\;\;\;\;\;\;\;\;\;\;\;\;\;\;\;\;\;\;\;\;\;\;\;\;\;\;\;\;\;\; + \beta \frac{\partial}{\partial \boldsymbol{r}} \left(\rho \boldsymbol{u} - [\rho \boldsymbol{u}]^{MD} \right) \bigg) \nonumber
\end{eqnarray}
\end{subequations}
where the intermolecular force and momentum is half from the normal MD dynamics with the other half made up by the average of the MD and continuum system for velocity and the remaining force being a result of the gradients in differences.
The gradient of the difference in density and pressure in the CFD and MD regions can be seen to apply a force driving the molecules

The flux terms $\phi_{\rho}$ and $\boldsymbol{\rho_u}$ are introduced in the derivation of \citet{Markesteijn_et_al_2014} to give a diffusion between the two phases.
These are apparently chosen as fluxes because this was found to give better behaviour than simply applying the direct difference between MD and continuum density and momenta.
Later work rewrites the diffusion in terms of surface fluxes.\citep{KOROTKIN2016446}
The equations are made conservative by ensuring the applied force to the MD system is equal and opposite to the continuum, where a fluctuating hydrodynamics model is used.
An assumption in this derivation is the external force on the system is equal to the divergence of the pressure tensor including the fluctuating component $F^C = \boldsymbol{\nabla} \cdot \left[\boldsymbol{P} +\boldsymbol{P}^{\prime} \right]$ for any system away from equilibrium. 
As a result, the molecular form of the pressure tensor does not appear in the equations.

\subsection*{An Example of Controlling Stress On 2 Surface in One Dimensional}

To understand how this works, consider a force which varies only in $y$, 
we can rewrite \eq{GLC_flux} as
\begin{align}
\Flin (y_i) =  - \FCV \bigg[  \overbrace{\Csurf{\!}_{y^+} N^+(y_i) -  \Csurf{\!}_{y^-} N^-(y_i)}^{\text{Linear}} \;\;\;\;\;\;\;\;\; \;\;\;\;\;\;\;\;\;
\nonumber \\
 - \underbrace{\eta \left[ y_i^2 - y_i(y^+ + y^-) + y^+ y^-\right]}_{\text{Quadratic}}  \bigg], 
\label{GLC_linear1D} 
\end{align}
so the forces applied on the top surface $\Csurf{\!}_{y^+}$ subtracts molecular surface pressure
and adds the CFD pressure value to drive the system to have pressure $A_y P_{y^+}$, while the bottom is driven 
toward $A_y P_{y^-}$ with a linear variation between them as shown schematically in Fig.~\ref{fgr:weights}.
A quadratic correction is then added, 
to ensure the total is as required to ensure the correct time evolution of momentum inside the volume.
Figure \ref{fgr:weights} shows an example of how we can use this to induce complex flow patterns, 
both elongation and rotational flow in two adjacent volumes,
while keeping momentum in both volumes the same \textit{i.e.} $d/dt \sum_{i=1}^N m_i \boldsymbol{r}_i \vartheta_i = 0$.


\balance


\bibliography{rsc} 
\bibliographystyle{rsc} 

\end{document}